\documentclass[traditabstract]{aa} 
\usepackage{graphicx}
\usepackage{subfigure}
\usepackage{lscape}
\usepackage{graphics}

\usepackage{color}
\usepackage{ulem} 
\usepackage{times}
\usepackage{float} %
\usepackage{txfonts}

\newcommand{\qh}{$Q({\rm{H^{0}}})$}

\newcommand{\Teff}{$T_{\rm{eff}}$}
\newcommand{\mTeff}{$\overline{T_{\rm{eff}}}$}
 \newcommand{\mU}{$\overline{U}$}

\newcommand{\Ha}{\ifmmode {\rm H}\alpha \else H$\alpha$\fi}
\newcommand{\Hb}{\ifmmode {\rm H}\beta \else H$\beta$\fi}

\newcommand{\hii}{H\,{\sc ii}}

\newcommand{\heii}{He\,{\sc ii}}
\newcommand{\Heii}{He\,{\sc ii}\,$\lambda$4686}

\newcommand{\Nii}{[N\,{\sc ii}]\,$\lambda$6584}

\newcommand{\nii}{[N\,{\sc ii}]}
\newcommand{\Oi}{[O\,{\sc i}]\,$\lambda$6300}
\newcommand{\oi}{[O\,{\sc i}]}
\newcommand{\Oii}{[O\,{\sc ii}]\,$\lambda$3727}

\newcommand{\oii}{[O\,{\sc ii}]}
\newcommand{\Oiii}{[O\,{\sc iii}]\,$\lambda$5007}
\newcommand{\Oiiit}{[O\,{\sc iii}]\,$\lambda$4363}
\newcommand{\oiii}{[O\,{\sc iii}]}

\newcommand{\Neiii}{[Ne\,{\sc iii}]\,$\lambda$3869}
\newcommand{\neiii}{[Ne\,{\sc iii}]}

\newcommand{\Nev}{[Ne\,{\sc v}]\,$\lambda$3426}
\newcommand{\Sii}{[S\,{\sc ii}]\,$\lambda$6716,\,$\lambda$6731}
\newcommand{\sii}{[S\,{\sc ii}]}
\newcommand{\Siii}{[S\,{\sc iii}]\,$\lambda$9069}

\newcommand{\siii}{[S\,{\sc iii}]}
\newcommand{\Ariii}{[Ar\,{\sc iii}]\,$\lambda$7135}
\newcommand{\ariii}{[Ar\,{\sc iii}]}

\newcommand{\ariv}{[Ar\,{\sc iv}]}

\newcommand{\rOiii}{[O\,{\sc iii}]\,$\lambda$4363/5007}

\newcommand{\rSii}{[S\,{\sc ii}]\,$\lambda$6731/6717}

\newcommand{\Hep}{He$^{+}$}

\newcommand{\Op}{O$^{+}$}
\newcommand{\Opp}{O$^{++}$}

\newcommand{\Nep}{Ne$^{+}$}
\newcommand{\Nepp}{Ne$^{++}$}
\newcommand{\Sp}{S$^{+}$}
\newcommand{\Spp}{S$^{++}$}
\newcommand{\Arpp}{Ar$^{++}$}
\newcommand{\Arppp}{Ar$^{+++}$}
\begin{document}
   \title{Excitation properties of galaxies with the highest \oiii/\oii\ ratios
}
   \subtitle{No evidence for massive escape of ionizing photons}

   \author{G. Stasi\'nska \inst{1}
\and
          Yu. Izotov \inst{1,2,3}
          \and
          C. Morisset \inst{4}
          \and 
           N. Guseva \inst{2,3}}
%
%
   \institute{LUTH, Observatoire de Meudon, F-92195 Meudon Cedex, France
         \and
                     Main Astronomical Observatory,
                     Ukrainian National Academy of Sciences,
                     Zabolotnoho 27, Kyiv 03680,  Ukraine
         \and
Max-Planck-Institut f\"ur Radioastronomie, 
Auf dem H\"ugel  69, 53121 Bonn, Germany            
         \and
                     Instituto de Astronom{\'\i}a, Universidad Nacional Aut\'onoma de M\'exico, Apdo. Postal 70264, M\'ex. D.F., 04510 M\'exico\\
             }
\offprints{grazyna.stasinska@obspm.fr}
   \date{Received  accepted }

  \abstract{
The possibility that star-forming galaxies may leak  ionizing photons is at the heart of many present-day studies that investigate the reionization of the Universe. 
We test this hypothesis  on local  blue compact dwarf galaxies of very high excitation. We assembled a sample  of such  galaxies by examining the spectra from Data Releases 7 and 10 of the Sloan Digital Sky Survey. We argue that reliable conclusions cannot be based on strong lines alone, and adopt a strategy that includes important weak lines such as \oi\ and the high-excitation  \heii\ and \ariv\ lines. Our analysis is based on  purely observational diagrams  and on a comparison of photoionization models with well-chosen emission-line ratio diagrams. We show that spectral energy distributions  from current stellar population synthesis models cannot account for all the observational constraints, which led us to mimick  several scenarios that could explain the data. These include the additional presence of  hard X-rays or of shocks. 
We find that only ionization-bounded models (or models with an escape fraction of ionizing photons lower than 10\%) are able to  simultaneously explain all the observational constraints.

   }
\titlerunning{Galaxies with the highest \oiii/\oii\ ratios}
\authorrunning{G. Stasi\'nska et al.}
   \maketitle


\section{Introduction}\label{intro}

By analyzing the entire Sloan Digital Sky Survey (SDSS-III) DR10 spectroscopic data base (Ahn et al. 2014), we found an 
appreciable number of compact star-forming galaxies with extremely high 
\oiii/\oii\ ratios\footnote{Throughout the paper, the notations \heii, \nii, \oi, \oii, \oiii, \neiii, \ariii, and \ariv\ will stand for \Heii, \Nii, \Oi, \Oii, \Oiii, \Neiii, \Ariii, and [Ar\,{\sc iv}]\,$\lambda$4740, respectively. }, exceeding values of 10 and reaching
up to 50. Hereafter these galaxies are called `extreme BCDs', standing for `extreme blue compact dwarf galaxies'.

\begin{table*}[h!t]
\centering
\caption{Atomic data used in our abundance determinations}
\label{tab:atomdat}
\begin{tabular}{lll}
\hline \hline 
Ion & Collision strengths & Radiative transition probabilities\\
\hline 
O\,{\sc ii} & Kisielius et al. (2009) & Zeippen (1982) \\
O\,{\sc iii} & Aggarwal \& Keenan (1999) & Galavis et al. (1997),  Storey \& Zeippen (2000) \\
N\,{\sc ii} & Tayal (2011) &   Galavis et al. (1997) \\
Ne\,{\sc iii} &  McLaughlin \& Bell 2000 & Galavis et al. (1997) \\
S\,{\sc ii} & Tayal \& Zatsarinny (2010) & Mendoza \& Zeippen (1982) \\
S\,{\sc iii} & Tayal \& Gupta (1999) & Froese Fischer et al. (2006) \\
Cl\,{\sc iii} & Mendoza (1983) &  Ramsbottom et al. (2001)\\
Ar\,{\sc iii} &  Munoz Burgos et al. (2009)  &    Munoz Burgos et al. (2009) \\
Ar\,{\sc iv} & Ramsbottom et al. (1997)  &   Mendoza \& Zeippen (1982)  \\
\hline
\end{tabular}
\end{table*}

Extreme BCDs
are local low-mass galaxies that are undergoing intense bursts of star formation.
They represent a subset of Green Peas  (Cardamone et al. 2009) when they are in the
redshift range 0.1 - 0.3, or of luminous compact galaxies (Izotov et al.
2011) in a wider redshift range 0.0 - 0.6. Therefore, similar to Green Peas and luminous compact galaxies,
extreme BCDs share  some of the properties of Ly$\alpha$ emitting
galaxies  and Lyman break galaxies     
at high-$z$: they are compact, have low-mass (10$^8$-10$^9$ $M_\odot$), 
low-metallicity (0.1-0.2 solar),  and
high specific star-formation rates (up to 10$^{-7}$ yr$^{-1}$)  
(Cardamone et al. 2009, Izotov et al. 2011).
Furthermore, unlike the typical local galaxies and like  Ly$\alpha$ emitting
galaxies  and Lyman break galaxies  (Nakajima et al. 2013, Shirazi et al. 2014, Shapley et al. 2014), extreme BCDs
are characterized by high excitation, as measured by the \oiii/\oii\ ratio.  
Thanks to these properties, extreme BCDs should be excellent local proxies for high-$z$  Ly$\alpha$ emitting
galaxies.

The high 
\oiii/\oii\ ratios can be attributed to several possible factors (Overzier et al. 2009, Kewley et al.
2013, Nakajima \& Ouchi 2013, Jaskot \& Oey 2013):
1) a low metallicity,
2) a high ionization parameter, 
3) a hard ionizing radiation field,
4) the presence of density-bounded H {\sc ii} regions.

This last hypothesis, if proven true, may have important cosmological implications in studies of the secondary 
ionization of the Universe, because Lyman continuum (LyC) emission freely escapes the density-bounded 
H {\sc ii} regions.

In this paper, we analyze the emission-line properties of local extreme BCDs  with the aim of finding out whether these high ratios are due to significant leakage of ionizing photons.

To construct our extreme BCD sample, we have thus searched the SDSS-III DR10 spectroscopic data base for  star-forming emission-line objects with the highest excitation, that is, those with the highest \oiii/\oii\ ratios  that show no sign  of an active galactic nucleus (AGN). We have merged this sample with a sample of high-excitation galaxies from SDSS DR7. 
For these extreme BCDs, in contrast to high-redshift objects, SDSS spectra provide a lot of information in addition to the intensities of the strongest lines, some of it being crucial for a proper interpretation of the excitation conditions.

This paper is organized as follows: In Sect. \ref{obsdata} we present the observational database, in Sect. \ref{trends} we discuss some purely observational trends that lead to a zero-order interpretation, in Sect. \ref{photo}
 we present the policy we  followed to interpret the observational data with the help of photoionization models, and in Sect. \ref{confront} 
 we compare various possible scenarios with the observations. The conclusions are summarized in the last section.

\section{The observational database}
\label{obsdata}

 We (Y. I.) searched  the  SDSS/BOSS DR10  spectroscopic database by eye to find star-forming galaxies with high \oiii/\oii\ Êratios ($\ga 5$). Since we were interested in star-forming galaxies, we removed objects with AGN-like spectra (i.e., altogether strong \oii\ and \oiii, broad lines, strong \oi, \nii, \sii, strong \heii, 
in some cases, red continuum). We ended up with a sample of 149 objects. All of them except one have \Oiiit\ detected.  In the following this sample is referred to as the DR10 sample.
 
 To this sample we added the sample of star-forming galaxies selected from SDSS DR7
(Abazajian et al. 2009) by two of us (Y. I. and N. G.) and described in Vale Asari et al. (2014). This sample is composed of all the star-forming galaxies in the SDSS DR7 in which the  \Oiiit\ line is seen and measured with an accuracy of at least 25 \%, and for which  visual inspection of the spectra and images showed that they  correspond to \textit{genuine} star-forming galaxies (i.e., excluding giant \hii\ regions in spiral galaxies and active galactic nuclei). Furthermore, we considered only
galaxies where \oii $\lambda$3727 is observed. 
With this criterion, all galaxies from the DR7 with $z \la 0.02$ are excluded.
In the following this sample is referred to as the DR7 sample. 

By combining our DR10 and DR7 samples, we obtained a sample of 778 objects,  of which 268  have \oiii/\oii\ ratios higher than 5, and 72 have  \oiii/\oii\ ratios higher than 10. The remaining objects are useful for the discussion because they allow us to consider a wider excitation sequence and ponder   the parameters leading to the highest observed values of \oiii/\oii.

The spectra from DR10 were analyzed with the IRAF software package\footnote{IRAF is distributed by the National Optical Astronomy Observatory, which is operated by the Association of Universities for Research in Astronomy, Inc., under cooperative agreement with the National Science Foundation.} in exactly the same way as described in Vale Asari et al. (2014) for the DR7 spectra. In brief, the emission-line fluxes were measured `manually' using the IRAF SPLOT routine. Correction of hydrogen line fluxes for underlying absorption and reddening correction of all lines were performed as described in that paper.

In both the DR7 and DR10 samples, the abundances of N, O, Ne, S, Cl, Ar, and Fe were derived from the intensities of the lines  
\Oii, \Oiii, \Nii, \Neiii, \Sii, \Siii, [Cl\,{\sc iii}]\,$\lambda$5518+5538, \Ariii, [Ar\,{\sc iv}]\,$\lambda$4740, and [Fe\,{\sc iii}]\,$\lambda$4988
with respect to H$\beta$. We used the temperature from \rOiii\ for doubly and triply ionized species  and a temperature derived from Eq. 14 of Izotov et al. (2006) for low-ionization species. The ionization correction factors used to obtain elemental abundances from ionic abundances were those described in Izotov et al. (2006). The atomic data involved in the computations are listed in Table \ref{tab:atomdat}.

The resulting abundance patterns for the DR10 and DR7 samples are presented in Appendix \ref{ab}. All the objects from our sample have low to moderate metallicities: 12 + log O/H in the range $7.5-8.5$. This is  a consequence of the fact that we imposed that the  \Oiiit\ line be present, implying that the electron temperature be sufficiently high, and thus that the metallicity be sufficiently low.

\section{Clues from observational trends}
\label{trends}

   \begin{figure}
   \centering
\includegraphics[scale=0.45, trim={0 0mm 0 0mm}, clip]{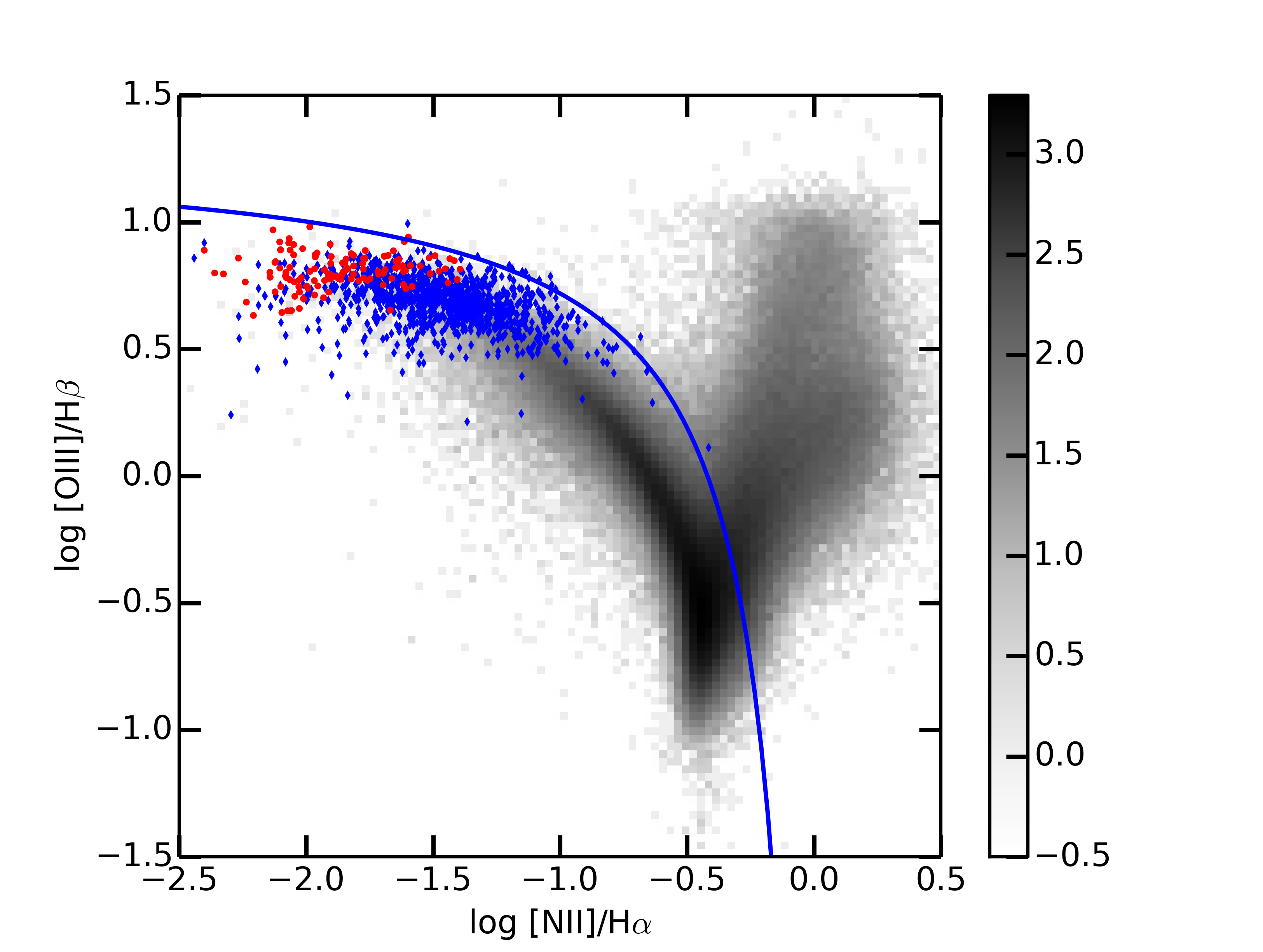}
      \caption{ BPT diagram for our DR10 objects (red circles) and those of  DR7  (blue diamonds). The background gray scale map represents all the SDSS DR7 galaxies with a signal-to-noise ratio of at least 3 in the four diagnostic lines \oiii, \Hb, \nii, and \Ha. The gray scale is in logarithm of the number of objects per pixel, as indicated in the bar on the right. The blue curve represents the Kauffmann et al. (2003) line generally used to distinguish star-forming galaxies from AGN hosts.}
         \label{figbpt}
   \end{figure}

In Fig. \ref{figbpt}, we represent the position of the DR7 and DR10 samples in the classical emission-line ratio diagram \oiii/\Hb\ vs \nii/\Ha\ (the famous BPT diagram from Baldwin et al. 1991). Throughout the paper, DR7 objects will be represented by blue symbols (diamonds in this figure) and DR10 objects by red symbols (circles in this figure).  The background gray-scale plot is a 2D-histogram   of all the galaxies in the 7th Data Release of the SDSS that have a signal-to-noise ratio larger than 3 for all the relevant lines. The gray scale at the right of the plot indicates the logarithm of the number of points in each pixel. We see that the number of points per gray pixel thus ranges from 1 to over 1000. The blue curve represents the empirical line drawn by Kauffmann et al. (2003) to distinguish star-forming galaxies from galaxies hosting an AGN.
We see that our selection criteria for the DR7 and DR10 indeed  samples selected star-forming galaxies, removing AGN hosts. We also see that they represent an extreme population of star-forming galaxies, with high values of \oiii/\Hb, and  \nii/\Ha\ ratios that reach extremely low values.

In the following, we present purely observational diagrams to investigate the relations between  \oiii/\oii\  and characteristic parameters of the emitting nebulae. These are the metallicity, expressed in units of 12 + log O/H and the  \Hb\ equivalent width, EW(\Hb), which measures the amount of ionizing photons 
produced by massive stars and absorbed by the gas with respect to the continuum at  \Hb, which is mainly produced  by the population of low-mass stars  in the galaxy (which, in turn, is a proxy for the galaxy stellar mass). 

Other parameters worthwhile considering are  $L$(\Hb), the  total \Hb\ luminosity  corrected for extinction, and the density $n_{\rm{e}}$, which can be derived from the \rSii\ line ratio. 

As is well known, two important parameters in the study of emission-line nebulae are the mean ionization parameter $\overline{U}$ (defined in Sect. \ref{grid}) and the mean effective temperature of the ionizing radiation field, \mTeff. 

In a first approximation, the abundance ratio of two adjacent ions of the same element, for example \Arppp/\Arpp,  is  given by the equation 
\begin{equation}
\label{ioni}
\frac{{\rm Ar}^{+++}}{{\rm Ar}^{++}} \propto \overline{U}  \frac{Q({\rm{Ar^{++}}})} {Q({\rm{H^{0}}})} \alpha(\rm{Ar}^{++}), 
\end{equation} 
where $Q({\rm{Ar^{++}}})$ is the number of photons above the ionization threshold of \Arpp, $Q({\rm{H^{0}}})$ is the number of hydrogen ionizing photons, and $\alpha(\rm{Ar}^{++})$  is the \Arpp\ recombination coefficient \footnote{Note, however, that the effect of charge transfer reactions with neutral hydrogen may modulate this relation, especially at low values of  
$\overline{U}$.} . 
The \Arppp/\Arpp\ ratio is thus a function of both $\overline{U}$ and of the mean effective temperature \mTeff\ of the ionizing radiation, since  $Q({\rm{Ar^{++}}})/Q({\rm{H^{0}}})$ is an increasing function of \mTeff\  (see Fig. \ref{fig:seds_BB} of Appendix \ref{seds}). 

The ionization parameter  $\overline{U}$  cannot be measured directly from observations, but, assuming a value for the volume-filling factor $\epsilon$, it can be obtained from the product $n_{\rm{e}}$ $L$(\Hb)   using Eq. \ref{Umeanfr03} below, since  $L$(\Hb) is a direct measure of the total number of hydrogen ionizing photons, \qh\ (in the hypothesis of no leakage and minimal absorption of LyC photons by dust). 

It has been proposed by V\'ilchez \& Pagel (1988) that  $\eta$, defined as (\Op/\Opp)/(\Sp/\Spp) is a good indicator of \mTeff. We actually  to use a ratio based on line intensities rather than on ionic fractions, since it is readily obtained from observations. We also prefer to use a ratio that increases with \mTeff. Thus, rather than the $\eta$ parameter, we use the (\oiii/\oii)/(\siii/\sii) ratio.
As shown in the bottom right panel of Fig. \ref{linerattiosvsT*} in Appendix \ref{didactic},  this ratio is far from being completely independent of $\overline{U}$, and is sensitive to temperature especially at low values of \Teff. A similar ratio is  (\ariv/\ariii)/(\oiii/\oii) , which is much less sensitive to  $\overline{U}$, but has a lower dynamic range as a function of \Teff\ (see  the bottom left panel of Fig. \ref{linerattiosvsT*}).

Finally, an important criterion to be considered when examining the question of LyC escape is the behavior of the \oi\ line. In density-bounded objects, this line is the first to be suppressed, or, at least, weakened.

\begin{figure}[!h]
\centering
\includegraphics[scale=0.9, trim={20 219mm 0 0mm}, clip]{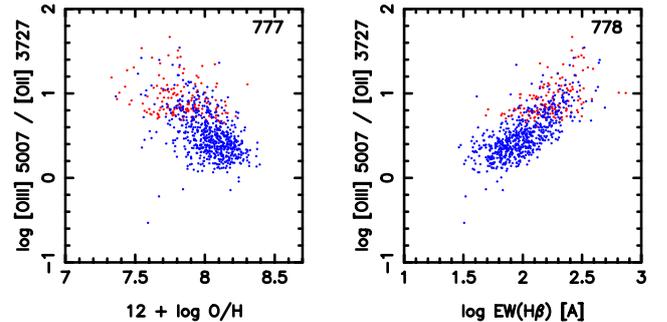}
\caption{The \oiii/\oii\ ratio as a function of the oxygen abundance 12 + log O/H (left panel) and as a function of the equivalent width of the \Hb\ line (right panel). DR10 objects are represented in red and those of DR7 in blue. The total number of objects in each panel is indicated in the top right corner.}
 \label{fig_O_EW}
\end{figure}

Figure \ref{fig_O_EW} shows the relation between \oiii/\oii\ and the oxygen abundance (left) and EW(\Hb) (right) for the DR7  and DR10  samples. Clearly, the highest values of  \oiii/\oii\ correspond to the lowest metallicities and to the highest values of EW(\Hb). Note that in the range of metallicities covered by our DR7 and DR10 samples, a decrease in metallicity from 12 + log O/H $=$ 8.5 to 7.4  produces an increase in  \oiii/\oii\ by about 0.3 dex if keeping the ionization parameter  equal (see Appendix \ref{didactic}, Fig. \ref{O3O2OsH}), while the observed variation is of about 1.3 dex. Thus, there is something special about the objects with the highest  \oiii/\oii, not only their low metallicity. The observed increase of  \oiii/\oii\ with  EW(\Hb) suggests an increase with \mTeff\ or perhaps an increase with $L(\Hb)$, the total $\Hb$ luminosity (since the continuum at $\Hb$ is mainly due to low-mass stellar populations and not to the ionizing stellar populations). In any case, it argues against increasing LyC photon leakage, since the latter would produce a drop in EW(\Hb). It must be noted that at the highest metallicities encountered in the studied galaxies, dust absorption reduces the  $\Hb$ luminosities  at the highest values of the ionization parameter (see Appendix  \ref{didactic}, Fig. \ref{HbQ}). But at the low metallicities where the highest \oiii/\oii\ values are found, the effects of 
dust absorption on EW(\Hb) are not important. 
\begin{figure}[!h]
\centering
\includegraphics[scale=0.9, trim={20 219mm 0 0mm}, clip]{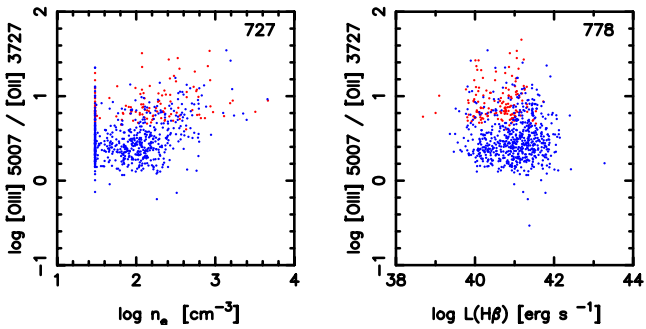}
\caption{The \oiii/\oii\ ratio as a function of the electron density as derived from the \rSii\ line ratio (left panel) and as a function of the total luminosity in the \Hb\ line, corrected for extinction. 
}
 \label{figLn}
\end{figure}

Figure \ref{figLn} shows the relation between \oiii/\oii\ and two other fundamental properties of the emitting regions: the electron density, $n_{\rm{e}}$, as derived from the \rSii\ line ratio (left), and  $L$(\Hb) (right). There is a slight tendency for  high values of  \oiii/\oii\ to occur at higher values of  $n_{\rm{e}}$ (the Pearson correlation coefficient $\mathcal{P}$ is 0.33). Note that the effect of collisional deexcitation at higher densities is only minor in this plot, since between log $n_{\rm{e}}$ $=2$ and 3.3, the emissivity ratio of \oiii\ and \oii\ increases by only 0.1 dex. There is no significant correlation between   \oiii/\oii\ and $L$(\Hb) ($\mathcal{P}=0.13$). 

\begin{figure}[!h]
\centering
\includegraphics[scale=0.9, trim={20 219mm 0 0mm}, clip]{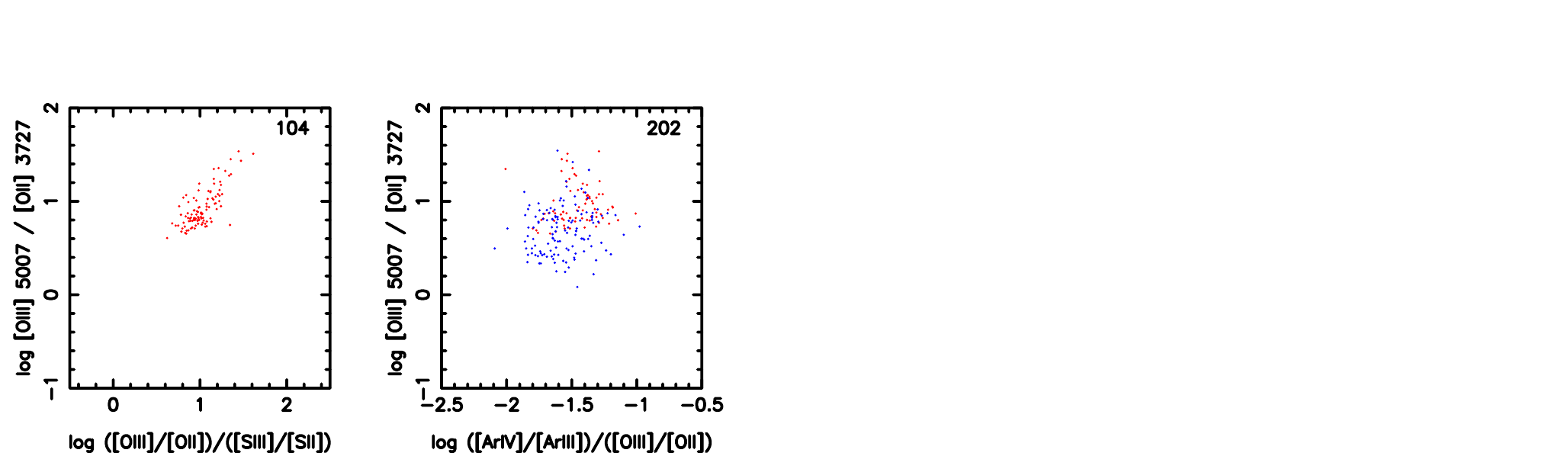}
\caption{The \oiii/\oii\ ratio vs (\oiii/\oii)/(\siii/\sii) (left panel) and  vs (\ariv/\ariii)/(\oiii/\oii) (right panel).
}
 \label{figT*}
\end{figure}

Figure \ref{figT*} shows the relation between \oiii/\oii\ and two parameters strongly related with the mean effective temperature of the ionizing radiation field, \mTeff: (\oiii/\oii)/(\siii/\sii) (left panel) and  (\ariv/\ariii)/(\oiii/\oii) (right panel). Both diagrams show a correlation, very strong for the first one ($\mathcal{P}  = 0.74$), weak but still significant for the second one ($\mathcal{P} = 0.21$). This suggests an increase of \oiii/\oii\  with \mTeff. Given that   (\oiii/\oii)/(\siii/\sii) is much more dependent on $\overline{U}$ than (\ariv/\ariii)/(\oiii/\oii) (see Appendix \ref{didactic},  Fig. \ref{linerattiosvsT*}), Fig. \ref{figT*} suggests that the trend observed in   \oiii/\oii\ expresses a trend both with increasing \mTeff\ and  increasing $\overline{U}$ in our sample of galaxies. 
{The increase in \mTeff\ might actually be a mere metallicity effect, since ionizing spectra of more metal-poor stellar populations are harder (see Fig. \ref{fig:SEDs_met} in Appendix \ref{seds}).

\begin{figure}[!h]
\centering
\includegraphics[scale=0.9, trim={20 219mm 0 0mm}, clip]{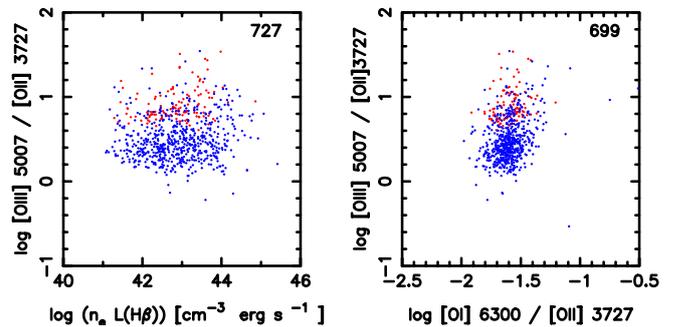}
\caption{The \oiii/\oii\ ratio as a function of the product $n_{\rm{e}}$ $L$(\Hb)   (left panel) and  as a function of \oi/\oii\ (right panel).
}
 \label{fig_U_esc}
\end{figure}

Figure \ref{fig_U_esc} (left) shows the relation between \oiii/\oii\ and  the product $n_{\rm{e}}$ $L$(\Hb), which as shown above, can be considered as a proxy for  $\overline{U}$ if the filling factor $\epsilon$ is constant. The fact that 
 \oiii/\oii\ is not seen to correlate with $n_{\rm{e}}$ $L$(\Hb)   suggests that the variations in  \oiii/\oii\  are due to an intrinsic increase of \mTeff\ rather than of  $\overline{U}$. Another option could be that  $\epsilon$  varies systematically from lower to higher values as    \oiii/\oii\ increases. We have no way to check this directly with the data at hand, but this is a  possibility. Figure \ref{fig_U_esc} (right) shows the relation between   \oiii/\oii\  and  \oi/\oii. The first thing to note is that most objects do show \oi\ emission, and that the \oi\ emission is not limited to objects with the lowest values of  \oiii/\oii. Actually, most of the objects with the highest values of  \oiii/\oii\ show  \oi/\oii\ ratios higher than 1\%. This suggests that the explanation for the highest values of  \oiii/\oii\ does not lie in LyC photon leakage. However, photoionization models are needed to confirm this interpretation. This is the subject of the next section.

\section{Photoionization model analysis}
\label{photo}

\subsection{The photoionization model grid}
\label{grid}

The models were constructed with the photoionization code {\sc Cloudy} version 13.03 (Ferland et al. 2013). 
The abundance ratios with respect to oxygen are defined as follows : 

O : N : Ne : S : Cl : Ar : Fe $=  1: 0.0513: 0.1854: 0.0219: 0.00029: 0.0048: 0.0148$, corresponding to the mean values of the logarithmic abundances derived for our merged DR7, DR10 sample (see Sect. \ref{obsdata} and Appendix \ref{ab}). Except for some of the models shown in Appendix \ref{didactic}, which explore the effects of varying O/H, we adopted for all the models a value of 12 + log O/H = 8, equal to the average of the measured abundances in our merged sample. The adopted helium abundance by number relative to hydrogen is 0.086, that of Mg and Si is 0.0095. 

The dust-to-gas abundance ratio is obtained by applying a factor $2/3 \times 10^{(2.21-y)}$ to the default value in Cloudy when using the `ism'  keyword. The expression for $y$ is obtained from Table 1 (Broken law, MW case) of R\'emy-Ruyer et al. (2014). The reduction factor of 2/3 is motivated by the study of Draine (2011).

For most models, the ionizing radiation field is given by the population synthesis code {\sc PopStar} (Moll\'a et al. 2009) for a Chabrier (2003) stellar initial mass function and at the appropriate metallicity, obtained by interpolation. The reference case corresponds to an age of 1 Myr.  Other ionizing radiation fields have also been considered, as described in the following sections. All the models assume a constant hydrogen density, but two geometries are considered. They are defined by two different values of $f_S= R_{\rm in}/R_S$, where $R_{\rm in}$ is the model inner radius  and $R_S$ is the Str\"omgren radius corresponding to $R_{\rm in}=0$. Models with  $f_S = 0.03$ are equivalent to a full sphere, models with $f_S = 3$ correspond to a  hollow spherical bubble.
The models are computed until the ratio of ionized hydrogen to total hydrogen density falls below 0.02.

In addition to the spectral energy distribution (SED) of the ionizing radiation, the gas chemical composition and density, one needs to define the luminosity of the ionizing source. It is more convenient, actually, to fix the volume-averaged ionization parameter,  $\overline{U}$ estimated in the following way. 
In a nebula, the ionization parameter $U(R)$ at a distance $R$ from the center is defined as 
\begin{equation}
\label{Udef}
U(R)=\frac{Q({\rm{H^{0}}})}{4 \pi R^2 n c}, 
\end{equation} 
 where $Q({\rm{H^{0}}})$ is the ionizing luminosity in photons s$^{-1}$, $n$ is the hydrogen density and $c$ the speed of light. The volume averaged ionization parameter, \mU, is defined by
 \begin{equation}
\label{Umean}
\overline{U}=\frac{\int U(R) dV}{\int dV},   
\end{equation}
 where the integration is performed over the volume of the ionized nebula. The radius of the ionized nebula is obtained by equating $Q({\rm{H^{0}}})$ with the total number of hydrogen recombinations to upper levels in the ionized volume. Assuming that the gas is uniformly distributed in the nebula with a constant density $n$ and a volume filling factor $\epsilon$, and assuming that  the case B recombination coefficient, $\alpha_{\rm B}$, is constant, we obtain
\begin{equation}
\label{Umeanfr03}
\overline{U}_{\rm input}= \left(\frac{3}{4 \pi}\right)^{1/3}  \frac{\alpha_B^{2/3}}{c} \left( Q(H^0) n \epsilon^2 \right)^{1/3}((1+f_S^3)^{1/3}-f_S).
\end{equation}

 Our series of models assume a density $n =100$ cm$^{-3}$ and take values of $\log \overline{U}_{\rm input}$ ranging from $-4$ to $-1$ by steps of 0.5 dex, which allows us to span the entire range of observed values of \oiii/\oii\footnote{In the insets of the figures we replaced $\log \overline{U}_{\rm input}$ by log $U$ for simplicity.}.  
 As we show below, slight differences occur in the predicted line ratios with the same value of $\overline{U}_{\rm input}$ for the two geometries considered.
 
 For each model, we created several density-bounded versions by trimming the outer parts and leaving only the inner parts corresponding to $f_{\Hb} =$ 20, 40, 60, 80, and 100\% of the value of the H$\beta$ luminosity obtained for the radiation-bounded model.

\subsection{How the models are used}
\label{approach}

In the present work we aim to determine the cause of the highest values of \oiii/\oii\ observed and, in particular, whether they can be attributed to LyC photon leakage. One of the important lines to consider is therefore \Oi, which is produced in the warm transition region between fully ionized gas and neutral gas. The diagnostic diagram we chose to consider is \oiii/\oii\ versus \oi/\oiii. In a first approximation, it is independent of O/H, which allows us to compare the models with 12 + log O/H $= 8$ with our entire observational sample. 

Since we suspect that the highest  \oiii/\oii\ might be linked to high values of \mTeff, it is of interest to use lines of ions with high ionization potentials as diagnostics. One such line is that of \ariv, since the ionization potential of  \Arpp\ is 40.6 eV, significantly higher than the ionization potential of \Op\ (35.1 eV). Another option would be the \neiii\ line, since the ionization potential of \Nep\ is also 40.6 eV, and, in addition, there is no charge transfer between \Nepp\ and \Nep, in contrast to what occurs between \Arppp\ and \Arpp. The advantage of \ariv\ is that it can be related to \ariii, and that the \ariv/\ariii\ ratio does not depend on abundances. The second diagnostic diagram we therefore chose to consider is \oiii/\oii\ versus \ariv/\ariii.

If we wish to understand the behavior of emission-line ratios in galaxies, we cannot put aside some lines, such as \heii, which, although weak, provide important diagnostics because they indicate  a very hard ionizing radiation field (above 54.4 eV). Their systematic presence in \hii\ galaxies has first been noted and commented  onby Garnett et al. (1991) and Stasi\'nska \& Izotov (2003), and was more thoroughly discussed by Shirazi \& Brinchmann (2012) and Jaskot \& Oey (2013). All the suggested explanations (Wolf-Rayet stars, shocks, X-ray binaries, stellar rotation effects on the effective temperature of massive stars, inaccurate predictions of stellar atmosphere models at high energies) can be disputed. In the present paper, our aim is not to propose an astrophysically valid explanation, so we  consider different scenarios to explain the observed \heii/\Hb\ Êratios and examine their effects on the remaining line ratios. We therefore choose as a third diagnostic diagram \oiii/\oii\ versus \heii/\Hb, and   retain photoionization models that predict \heii/\Hb $\simeq 0.01$, which is representative of the bulk of the objects in our sample that show \heii\ Êemission (the \heii\ line is not detected in about one third of the objects of our sample, but in at least part of these cases this absence can be attributed to a too low signal-to-noise ratio).

\section{Comparing models with observations}
\label{confront}

\subsection{Models using classical stellar population synthesis results}
\label{classical}

\subsubsection{The reference model: {\sc PopStar} with an age of 1 Myr}
\label{popstar1}

   \begin{figure*}[!htb] 
   \centering 
  \makebox[\textwidth][c]{\includegraphics[scale=0.45, trim={10 00mm 0 0mm}, clip]{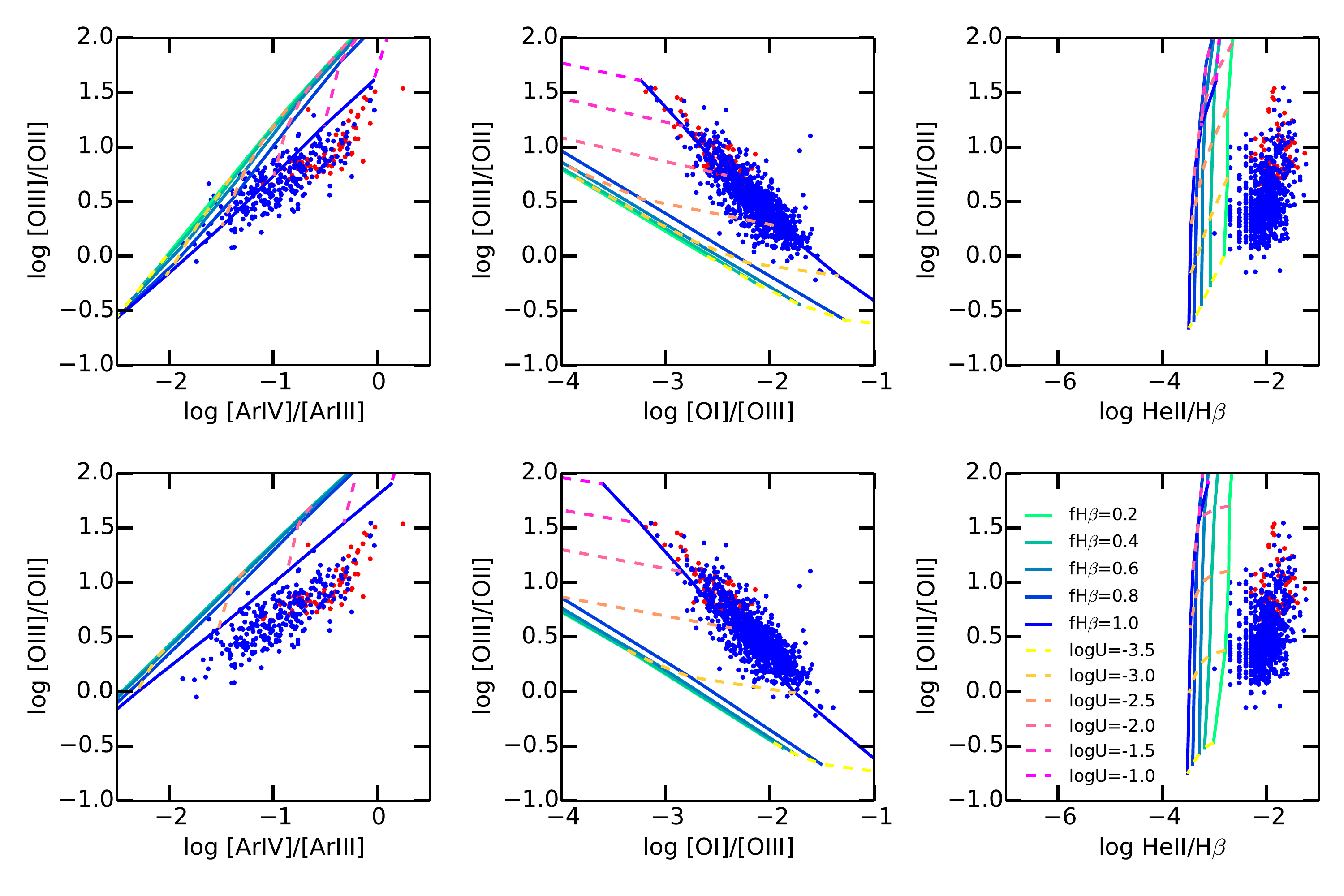}
}
      \caption{The \oiii/\oii\ versus \ariv/\ariii\ (left),  \oiii/\oii\ versus \oi/\oiii\ (middle) and 
 \oiii/\oii\ versus \heii/\Hb\ (right) diagnostic diagrams. The red and blue points represent the DR10 and DR7 objects respectively. The curves represent our grid of models computed with the SED from {\sc PopStar} models at an age of 1 Myr and  with a metallicity 12 + log O/H $= 8$. The dashed lines link models with the same values of $\overline{U}_{\rm input}$,  the continuous lines link models with the same value of  the trimming parameter $f_{\Hb}$, the ionization-bounded models corresponding to $f_{\Hb}=1$. The key for the colors is given in the bottom right panel. The upper row of panels corresponds to our filled sphere models ($f_S = 0.03$), while the lower row corresponds to our bubble models ($f_S = 3$).
 }
         \label{figpopstar1}
   \end{figure*}


Figure \ref{figpopstar1} shows the distribution of the observational points in our three diagnostic diagrams: \oiii/\oii\ versus \ariv/\ariii\ (left),  \oiii/\oii\ versus \oi/\oiii\ (middle), and 
 \oiii/\oii\ versus \heii/\Hb\ (right). Superimposed is our grid of photoionization models computed with the SED from {\sc PopStar} models at an age of 1 Myr and with an oxygen abundance 12 + log O/H $= 8$. The dashed lines link models with the same values of $\overline{U}_{\rm input}$, while the continuous lines link models with the same value of  the trimming parameter $f_{\Hb}$. The upper row of panels correspond to our filled sphere models ($f_S = 0.03$), while the lower row corresponds to our bubble models ($f_S = 3$). It can be seen that these models fall very short of reproducing the observed  \heii/\Hb\ ratios, even if considering density-bounded models. The SED from the {\sc PopStar} models  at an age of 1 Myr simply does not provide enough \Hep\ ionizing photons. The observational points fall well on the ionization-bounded models line in the   \oiii/\oii\ versus \oi/\oiii\ diagram\footnote{It is interesting to note that, in the  \oiii/\oii\ versus \oi/\oiii\ plot, the curves for  $f_{\Hb} <1 $ almost overlap, with only the $f_{\Hb} =1 $ curve standing out. This is because in that case the \oi\ emission is dominated by the ionization front, while in the density-bounded cases it is produced by the traces of neutral hydrogen  in the whole   \hii\ region.}, both for the filled and hollow sphere models. However, this cannot be considered as proving that the objects are all ionization-bounded, since the models fail to reproduce the \heii/\Hb\ and the \ariv/\ariii\ line ratios  (especially for the hollow sphere models). We note that the highest  observed \oiii/\oii\ ratios correspond to the highest values of  $\overline{U}_{\rm input}$ considered in our models. We also note that the hollow bubble models seem to provide a poorer fit to the observations in the \oiii/\oii\ versus \ariv/\ariii\ diagram than the filled sphere models, but this question cannot be fully addressed until the SED problem is solved (see below).

\subsubsection{Other stellar populations}
\label{otherseds}

In Fig. \ref{fig:seds_SEDs_1234_1} of Appendix \ref{seds} we show the SEDs obtained at an age of 1 Myr with the same stellar initial mass function, but using the different assumptions for the stellar evolution models proposed by {\sc starburst99} (Leitherer at al. 1999): Geneva tracks with standard mass loss, Geneva tracks with high mass loss, original Padova tracks, and Padova tracks with AGB stars. All the SEDs have been obtained with the stellar atmosphere grid of  Smith et al. (2002) based on atmosphere models of Pauldrach et al. (2001) and Hillier \& Miller (1998), similarly to the {\sc PopStar} models. From this figure, it is clear that at an age of 1 Myr, all these SEDs are equivalent and will lead to almost identical photoionization models.

Another question to consider is the age of the ionizing stellar populations. Fig. \ref{fig:seds_age} of Appendix \ref{seds} shows the SEDs obtained by {\sc PopStar} for several ages. We see that, for an age of 2 Myr, the SED is almost identical to that at 1 Myr, and that for 3 and 5 Myr, the SEDs are softer. This implies that they will lead to the same problems regarding \ariv/\ariii\ and \heii/\Hb. However, at an age of 4 Myr, which is when massive Wolf-Rayet stars begin to come into play, the SED is much harder.

   \begin{figure*}[!htb] 
   \centering 
  \makebox[\textwidth][c]{\includegraphics[scale=0.45, trim={10 00mm 0 0mm}, clip]{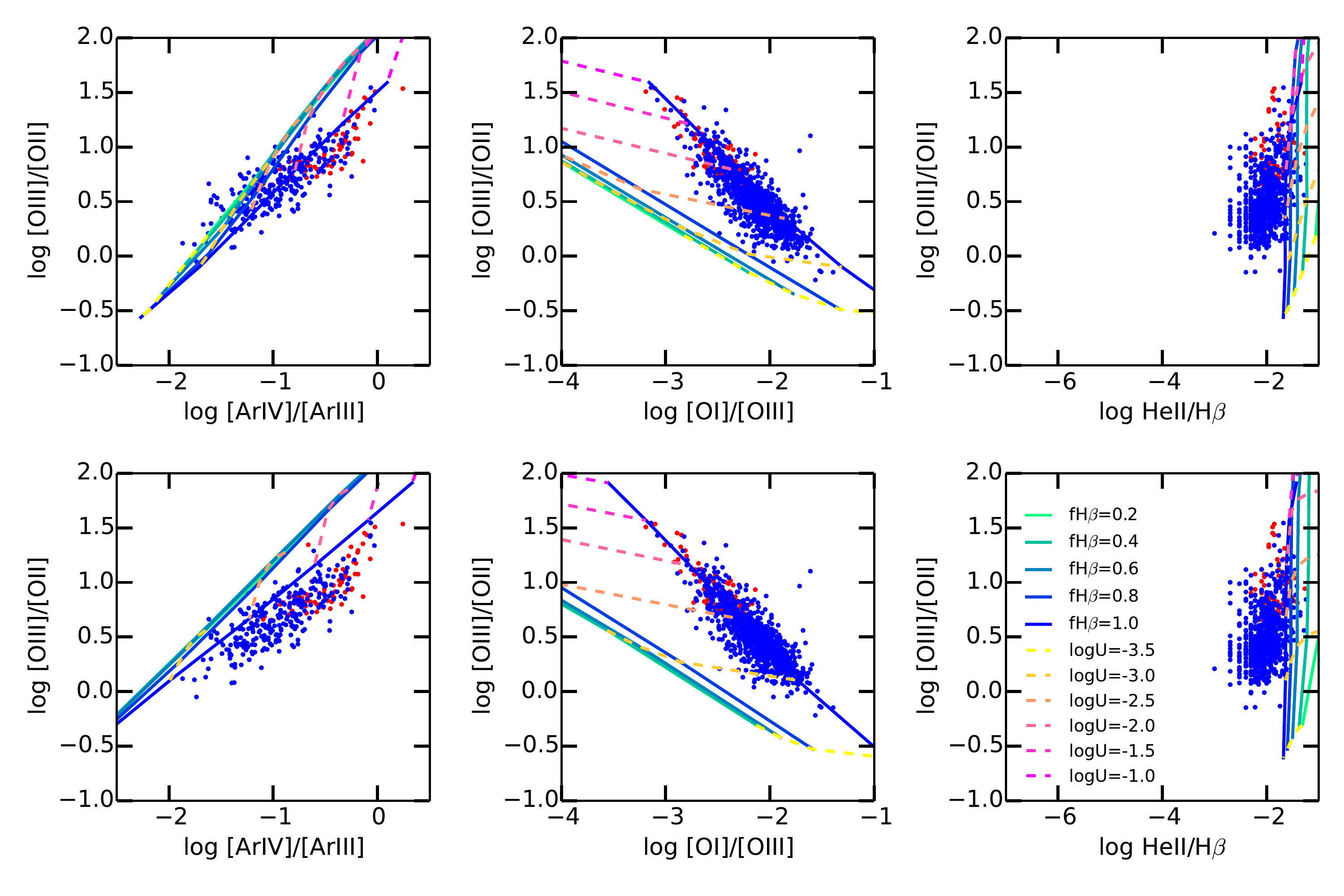}
}
      \caption{Same as Fig. \ref{figpopstar1}, but for models at an age of 4 Myr.
 }
         \label{figpopstar4}
   \end{figure*}

Figure \ref{figpopstar4} shows the results of photoionization models computed with  {\sc PopStar} for an age of 4 Myr, using the same presentation as Fig.  \ref{figpopstar1}. We see that the   \oiii/\oii\ versus \oi/\oiii\ Êdiagram is again compatible with ionization-bounded models, but that the SED at this age is even too hard to account for the bulk of the observed \heii/\Hb\ ratios. One then might think that the  relevant ages for reproducing the \heii/\Hb\ ratios is larger than 3 Myr and somewhere in the vicinity of 4 Myr. 

However,  this cannot be the explanation for all the objects that show \heii\ emission. The \heii\ line has been observed in 488 objects out of the 778 objects in our sample. In a first approximation, in absence of leakage or dust absorption of ionizing photons, and assuming that all the photons with energies higher than 54.4 eV are absorbed by \Hep\ (which is roughly correct except at low values of the ionization parameter, see Stasi\'nska \& Tylenda 1986), we have 
\begin{equation}
\label{eq:HesH}
 \rm{He} {\sc II}/\rm{H}\beta = A Q(\rm{He}^{+}) / Q(\rm{H}^0),
\end{equation} 
 where $A$ is the ratio of emissivities of the \heii\ and \Hb\ lines divided by the ratio of the case B recombination coefficients $\alpha_{\rm B}(\rm{He}^{+}) /\alpha_{\rm B}(\rm{H}^{0})$.  The coefficient $A$ depends only weakly on the electron temperature and is equal to 1.74 at 20,000~K. The left panel of Fig. \ref{HeWmod} shows the values of $Q(\rm{He}^{+}) / Q(\rm{H}^0)$ predicted by {\sc PopStar} models of metallicities $Z=0.004, 0.008$ and 0.02 (which correspond to 12 + log O/H $= $7.7, 8  and 8.4) as a function of age. We see that $Q(\rm{He}^{+}) / Q(\rm{H}^0)$ leads to a detectable \heii\ only during an age range of  3.9--4.25, 3.4--5.4, and 4.0--5.25 Myr,  respectively, for the three metallicities considered, if we assume detectability at \heii/\Hb\ $\ge 10^{-3}$. The right panel of Fig. \ref{HeWmod} shows the values of \heii/\Hb\ as a function of  EW(\Hb) for the same  {\sc PopStar} models, taking Eq. \ref{eq:HesH} to estimate \heii/\Hb. This figure can be directly compared to the observations, shown in the left panel of Fig. \ref{HeWobs}, having in mind the observed distribution of metallicities as a function of  EW(\Hb), shown in the right panel of Fig. \ref{HeWobs}. We see that the models do not predict any \heii\ emission for  EW(\Hb) above 200 \AA, while many of our objects (and especially some of those with the highest values of \oiii/\oii\ (see Fig. \ref{fig_O_EW}) show higher values of EW(\Hb). We must note, in addition, that the computed values of EW(\Hb) correspond to an instantaneous starburst, while in real objects, part of the continuum comes from old stellar populations that do not contribute to the ionization, so that the observed values of EW(\Hb) are lower than the theoretical values for the most recent starburst. This means that, even at values of  EW(\Hb) lower than 200 \AA, it is likely that in many of the  objects the observed \heii\ line is not due to ionization by a simple stellar population. Considering the possibility of extended starbursts rather than instantaneous ones does not help since it dilutes the \heii\ emission simultaneously with the \Hb\ emission. A similar discussion can be found in Shirazi \& Brinchman (2012) who, in addition, found that 40 per cent of the galaxies in their sample do not have Wolf-Rayet features in their spectra despite showing strong nebular \heii\ emission.

The above discussion of the production of nebular \heii\ emission by stellar populations of about 4 Myr relies on {\sc PopStar} models. Fig. \ref{fig:seds_SEDs_1234_4} of Appendix \ref{seds} shows the SEDs obtained at 4 Myr for the same stellar synthesis models as Fig. \ref{fig:seds_SEDs_1234_1}. We see that, at this age, the SEDs differ significantly owing to the different stellar evolutionary tracks involved. However, it can be intuited that the above discussion remains qualitatively valid and that the   \heii\ emission observed in the objects of our sample cannot be in majority caused by Wolf-Rayet stars.

In the next subsections, we explore various scenarios based on modified ionizing SEDs that might better explain the observed trends in our diagnostic diagrams. 


\subsection{Models using blackbodies as proxies for a harder stellar radiation field}
\label{bb}

   \begin{figure} 
       \begin{minipage}[t]{4.5cm}
   \centering
{\includegraphics[scale=0.33, trim={10 0mm 0 0mm}, clip]{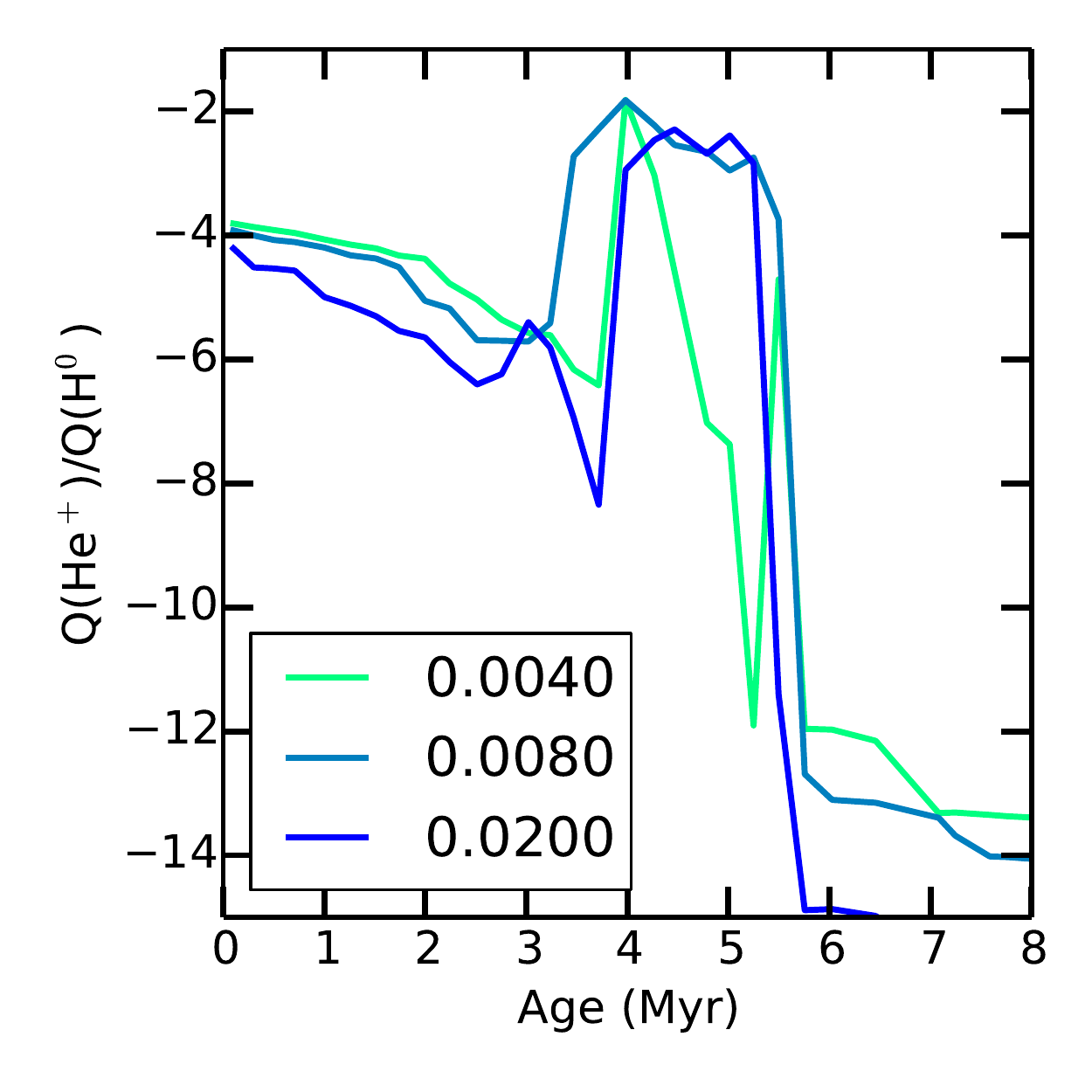}
}
       \end{minipage}
    \begin{minipage}[t]{4cm}
   \centering
{\includegraphics[scale=0.33, trim={0 0mm 0 0mm}, clip]{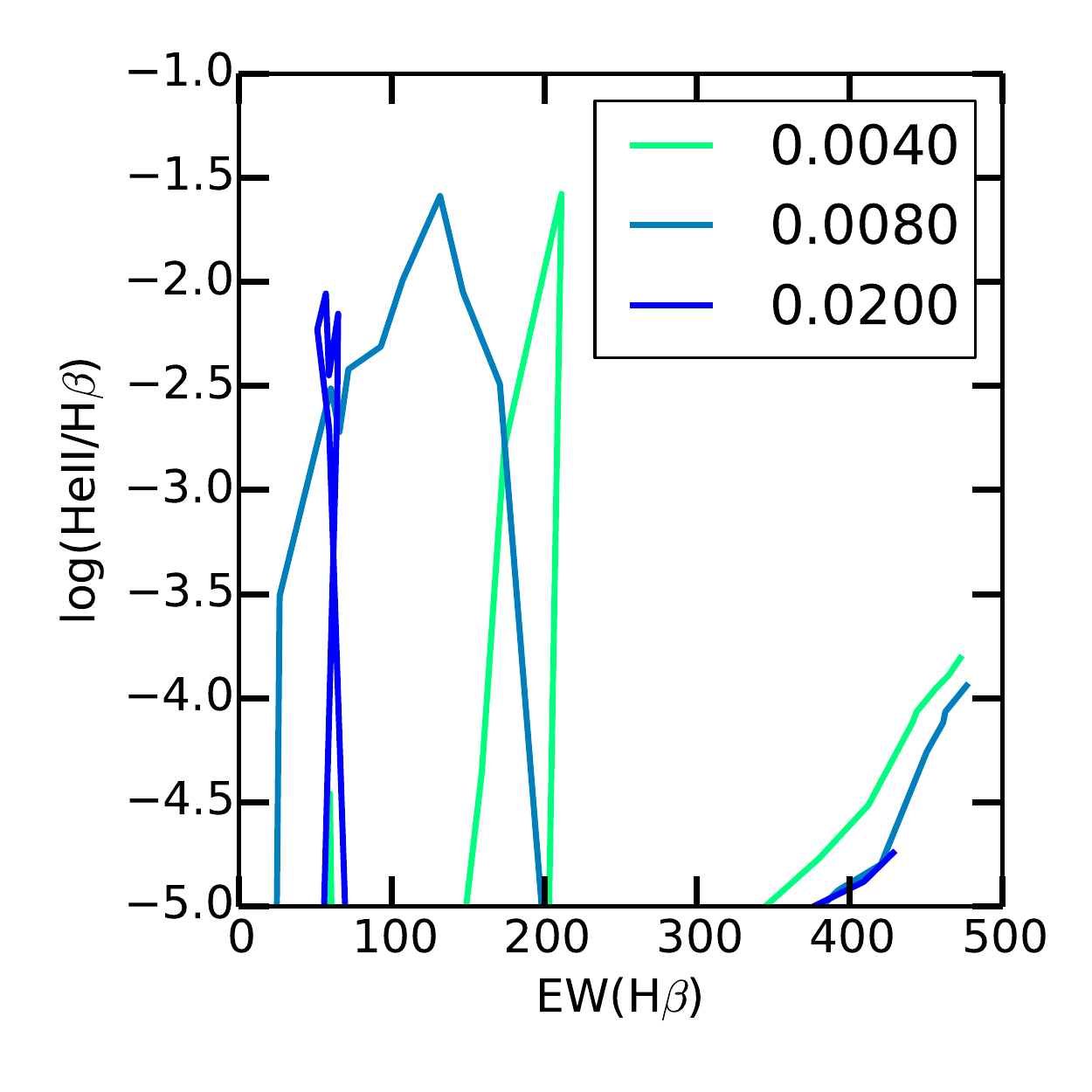}
}
       \end{minipage}
       \caption{Left: $Q(\rm{He}^{+}) / Q(\rm{H}^0)$ as a function of age for {\sc PopStar} SEDs with metallicities $Z=$0.004, 0.008, and 0.02 as indicated in the inset. Right: log \heii/\Hb\ versus EW(\Hb), where \heii/\Hb\ is taken equal to $1.74 Q(\rm{He}^{+} / Q(\rm{H}^0)$, for the same  {\sc PopStar} models. }
         \label{HeWmod}
   \end{figure}

   \begin{figure} 
   \centering 
\includegraphics[scale=0.9, trim={20 219mm 0 0mm}, clip]{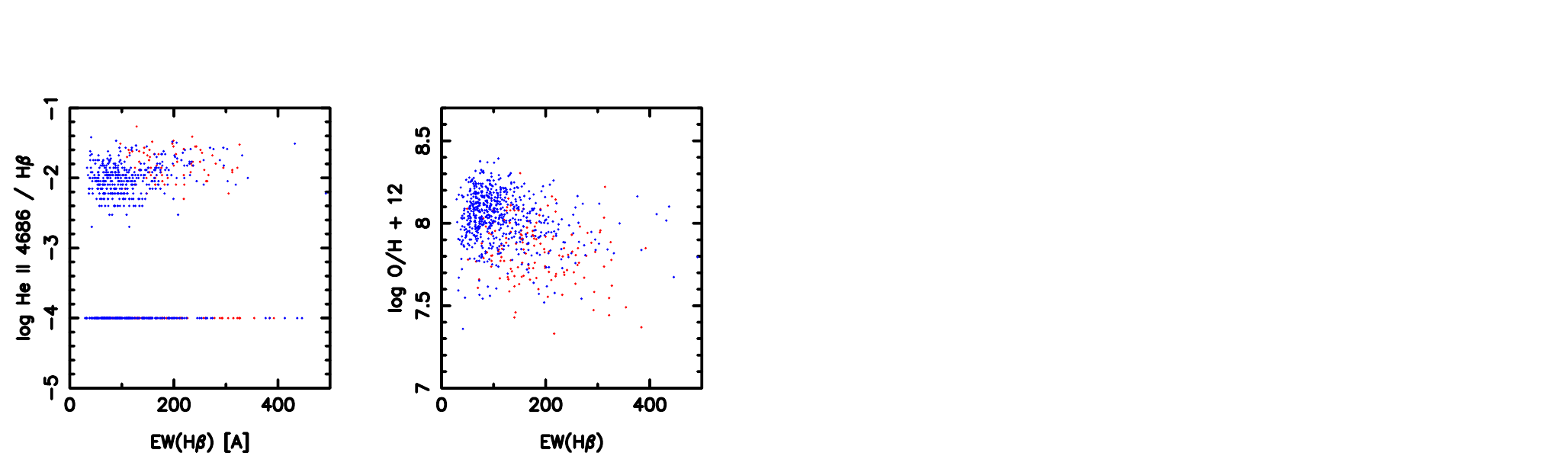}
       \caption{Left: observed values of log \heii/\Hb\ versus EW(\Hb) (for objects with no \heii\ detection log \heii/\Hb\ has been set to -4); Right: observed values of 12 + log O/H versus EW(\Hb).}
         \label{HeWobs}
   \end{figure}

   \begin{figure*} [!htb]
   \centering
  \makebox[\textwidth][c]{\includegraphics[scale=0.45, trim={10 00mm 0 0mm}, clip]{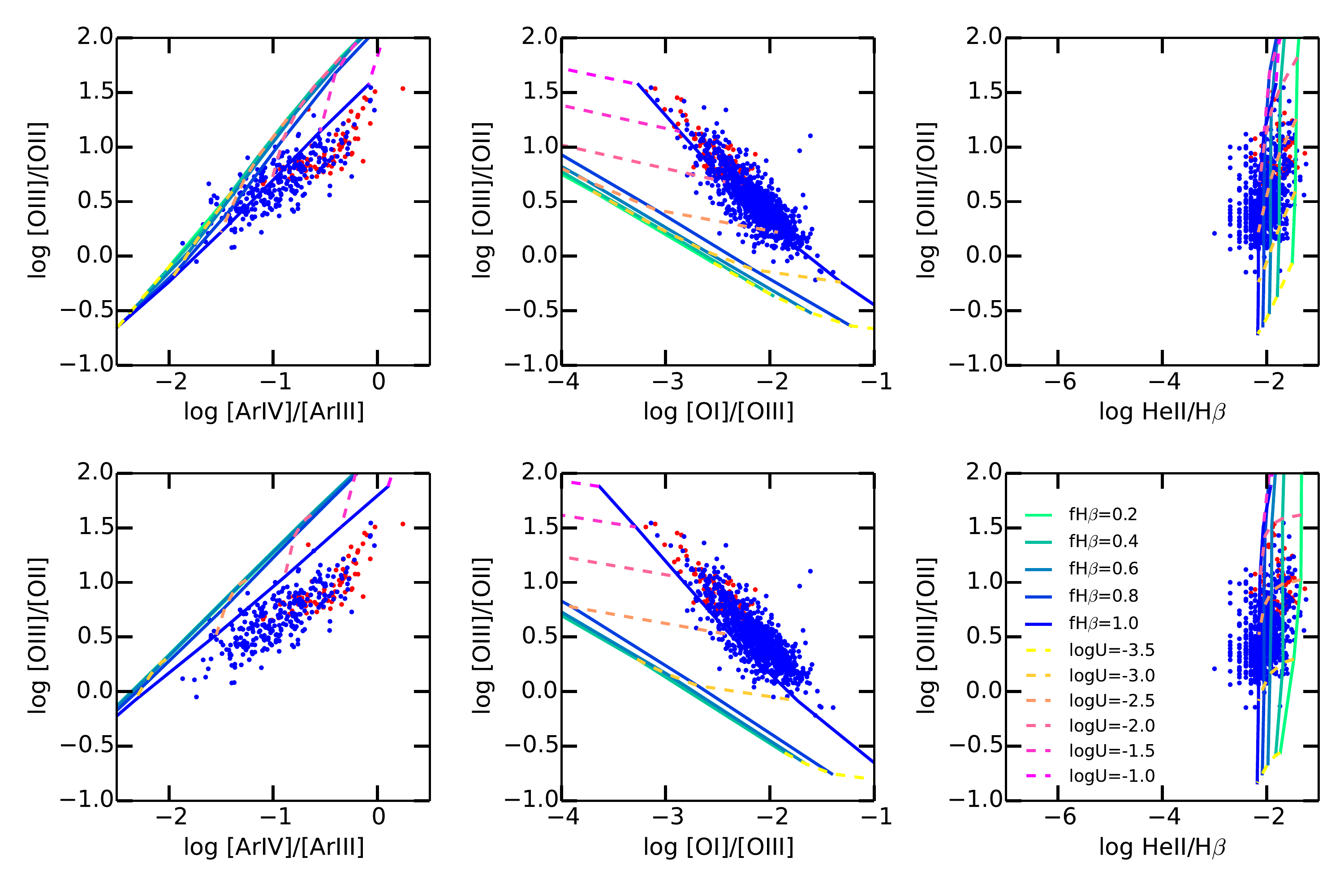}
}
      \caption{Same as Fig. \ref{figpopstar1}, but with models computed with a blackbody ionizing source at a temperature of 60,000 K. 
 }

         \label{figbb}
   \end{figure*}

Here we consider simple  blackbodies as proxies for the SEDs of the ionizing stellar populations. We find that a blackbody with \Teff$=$60,000 K leads to a value of $\sim$ 0.01 for \heii/\Hb\ in ionization-bounded models. Figure \ref{figbb} shows the results of models built with this ionizing radiation field for our three diagnostic diagrams (i.e., it is the analog of Fig. \ref{figpopstar1} for this SED). We see that the values of  \oi/\oiii\ fall on the ionization-bounded line, but that, again, the predicted \ariv/\ariii\ line ratios are too low. Considering slightly cooler blackbodies and density-bounded models, while acceptable for the \oiii/\oii\ vs  \heii/\Hb\ diagram would be incompatible with the \oiii/\oii\ vs  \ariv/\ariii\   diagram, since for a given value of \oiii/\oii\ the models lead to lower values of  \ariv/\ariii\  as \Teff\ decreases, as shown in Fig. \ref{o3o2a4a3}.

We remark that the \oiii/\oii\ vs  \ariv/\ariii\ diagram demands a harder SED than that of a blackbody at 60,000 K, especially at the highest values of \oiii/\oii. From  Fig. \ref{o3o2a4a3} one can estimate that the SED above 40.8 eV should provide roughly as many  photons as a blackbody at 90,000 K or more. Current synthetic stellar population models do not consider stellar atmospheres with such a hard radiation field. However, it is to be noted that some of the models for very massive stars at low metallicities computed with an appropriate treatment of radiation-driven winds (Kudritzki 2002) do provide the amount of photons needed to reproduce both the \oiii/\oii\ vs  \ariv/\ariii\  and the  \oiii/\oii\ vs  \heii/\Hb\ diagrams. Perhaps using more realistic model atmospheres to construct SEDs from stellar population models could help.

   \begin{figure}[!h] 
   \centering
\includegraphics[scale=0.35, clip]{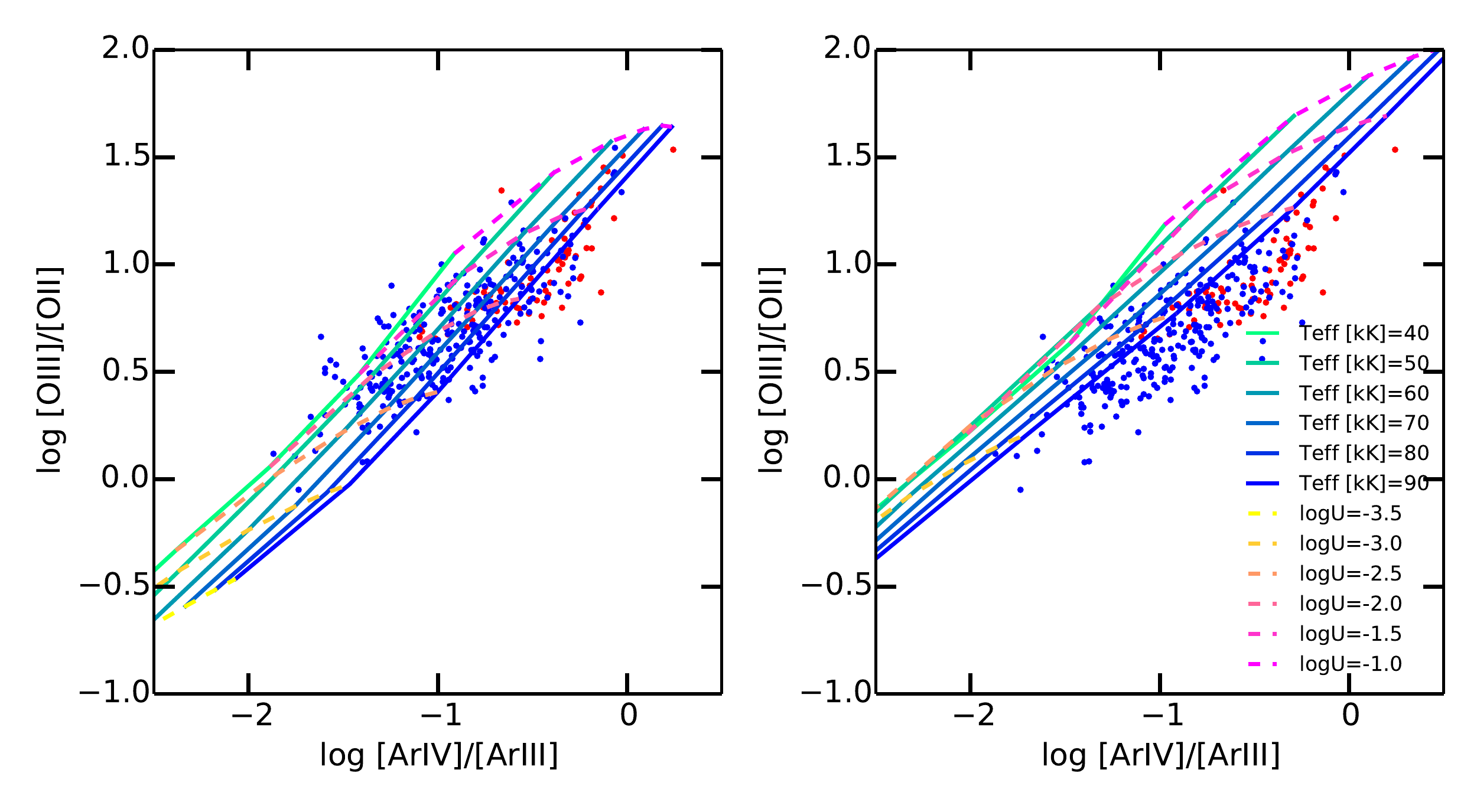}
      \caption{The \oiii/\oii\ vs \ariv/\ariii\ diagram for ionization-bounded photoionization models with a blackbody ionization source (12 + log O/H $=8$). Top panel: models for filled spheres; bottom panel: models for spherical bubbles.  Continuous lines correspond to different values of the effective temperature, while dashed lines correspond to different values of the ionization parameter. The color coding is indicated in the plot. The positions of the observational points for our samples (red for DR10, blue for DR7) are shown for comparison. This diagram shows that for a given value of \oiii/\oii\ the value of \ariv/\ariii\ increases with \Teff.
      }
         \label{o3o2a4a3}
   \end{figure}


\subsection{Models using classical stellar populations and an additional bremsstrahlung}
\label{brem}

   \begin{figure*} 
   \centering
  \makebox[\textwidth][c]{\includegraphics[scale=0.45, trim={10 00mm 0 0mm}, clip]{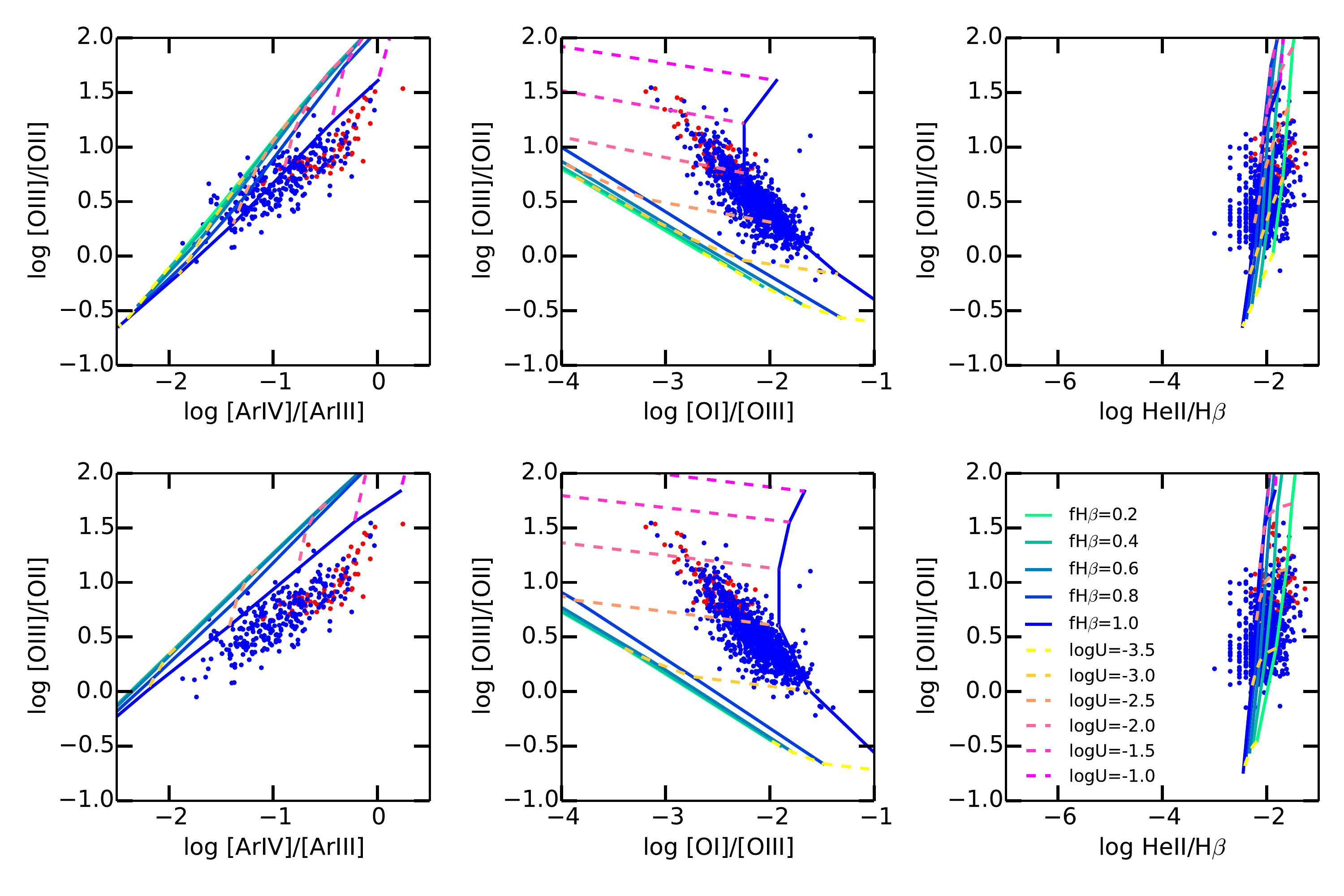}
}
      \caption{Same as Fig. \ref{figpopstar1}, but with models in which a bremsstrahlung at 10$^8$K is added to the stellar ionizing radiation from the {\sc PopStar} models corresponding to an age of 1 Myr, in such a proportion that 1\% of the ionizing photons come from the bremsstrahlung. 
      } 
         \label{figbr}
   \end{figure*} 

Another possibility to explain the observed values of  \heii/\Hb\ might be the presence of massive X-ray binaries (Van Bever \& Vanbeveren 2007) or of shocks in the atmospheres of stars leading to production of X-rays (Pauldrach et al. 2001), or else X-ray emission from protostars (Bonito et al. 2010).  We mimicked all these possible scenarios by adding a bremsstrahlung component with a temperature of 10$^8$K to the {\sc PopStar} models of Sect. \ref{popstar1}. This temperature is typical of the values inferred from a detailed analysis of  A 0535+26, one of the best-studied massive X-ray binary (Rothschild et al 2013). We adjusted the contribution from the bremsstrahlung so as to obtain  \heii/\Hb$\simeq 0.01$ in the resulting photoionization models. This implies that 1\% of the ionizing photons  come from this bremsstrahlung component of the spectrum.    The resulting SED is shown in Fig. \ref{fig:seds_AGNbrem} of Appendix \ref{seds}. The results of the photoionization models in our diagnostic diagrams are shown in Fig. \ref{figbr} (to be compared to Fig. \ref{figpopstar1}). We see that, at the highest values of \mU\ the ionization models slightly deviate from the observations in the \oiii/\oii\ vs  \oi/\oiii\ plot, but that density-bounded models reproducing the observations have a  LyC photon escape fraction of 10\% or lower. 
We note, however, that in the \oiii/\oii\ vs  \ariv/\ariii\ plot the observational points tend to lie below the models, meaning that the ionizing source does not provide enough photons at energies slightly above 40.6 eV. 


\subsection{Models using classical stellar populations and adding the effects of shocks}
\label{shocks}

   \begin{figure*} [!htb]
   \centering
  \makebox[\textwidth][c]{\includegraphics[scale=0.5, trim={10 5mm 0 0mm}, clip]{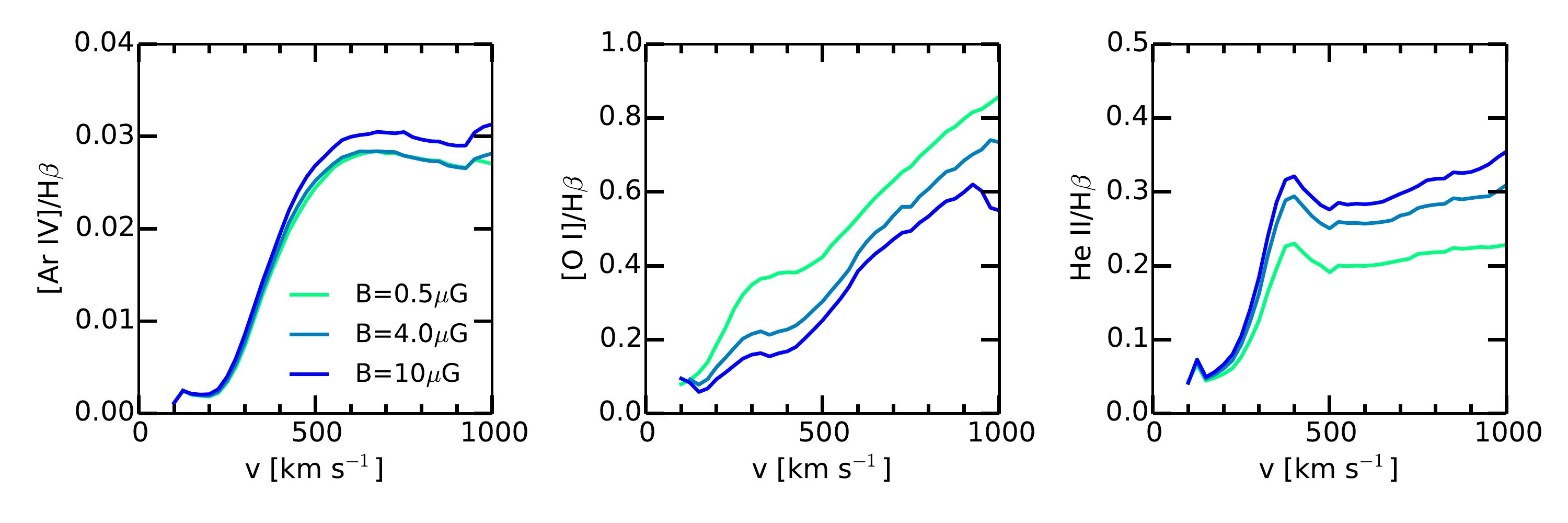}
}
      \caption{The line ratios \heii/\Hb\ (left panel), \oi/\Hb\ (middle panel), and \ariv/\Hb\ (right panel) as a function of shock velocity for the shock models from Allen et al. (2008) (see text). The different curves correspond to different values of the magnetic field, with the color code indicated in the left panel. 
      } 
         \label{shock}
   \end{figure*}

Another option to consider is the effects of shocks either produced by supernovae from previous generations of stars or from random motions of interstellar clouds. Such shocks could easily produce \heii\ lines  and were even advocated to explain the \Nev\ emission observed in some star-forming galaxies as predicted by photoionization models (Izotov et al. 2012). We adopted the strategy of Izotov et al. (2012) and Jaskot \& Oey (2013) to add the contribution of shocks to the emission-line properties of star-forming galaxies. We used the grid of radiative shock models from Allen et al. (2008)  and extracted from it the models constructed with the SMC abundance set because this chemical composition is the most appropriate for our case among the available compositions.

Figure \ref{shock} shows the variations of  \ariv/\Hb, \oi/\Hb, and \heii/\Hb\  as a function of the shock velocity for three different values of the magnetic field.  
The photoionization models and shock models are combined in the following way:
\begin{equation}
\label{ }
I_{\rm c}/H\beta_{\rm c} = \frac{(I_{\rm ph}/H\beta_{\rm ph} + I_{\rm sh}/H\beta_{\rm sh} \times C )}{(1 + C)}, 
\end{equation}
where $I/H\beta$ stands for the intensity of a line relative to \Hb, the subscripts `c', `ph', and `sh' stand for `combined', `photoionization', and `shock', respectively, and $C$ is the adopted ratio of $L(H\beta)_{\rm sh}/L(H\beta)_{\rm ph}$.

   \begin{figure*} 
   \centering
  \makebox[\textwidth][c]{\includegraphics[scale=0.45, trim={10 5mm 0 0mm}, clip]{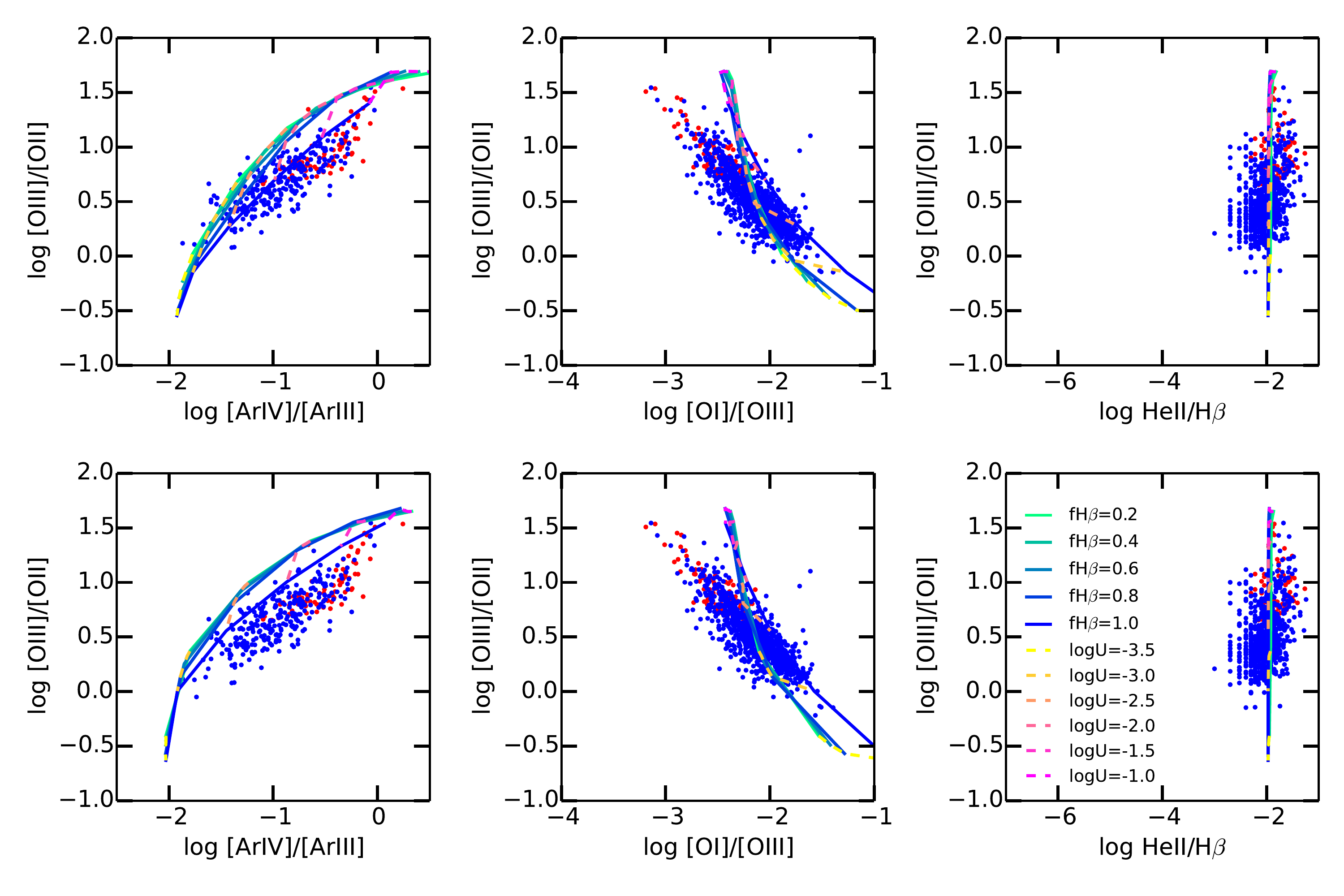}
}
      \caption{Same as Fig. \ref{figpopstar1}, but with combined photoionization and shock models. The photoionization models use the energy distribution of {\sc PopStar} at an age of 1 Myr. The shock models are extracted from the shock-model library of Allen et al. (2008) for a magnetic field of 0.5 $\mu$G and have a velocity of 300 km s$^{-1}$. They are added in such a proportion to obtain  \heii/\Hb$= 0.01$  for the combined model (see text for more details). }
         \label{fig4c}
   \end{figure*}

In Fig. \ref{fig4c} we show the combination of the models using {\sc PopStar} with a shock model with a velocity of 
300 km s$^{-1}$  and  a magnetic field strength of 0.5 $\mu$G. This value of the magnetic field minimizes the contribution to \heii/\Hb\ and maximizes the contribution of \oi/\Hb. The value of $C$ to obtain  \heii/\Hb$= 0.01$   is 0.09. This figure shows that, at the highest values of  \oiii/\oii\ the combined models produce an  \oi/\Hb\ ratio significantly higher than observed, regardless of the value of the trimming parameter $f_{\Hb}$. The reason is that the  \oi\ emission in a combined model that explains the  \heii/\Hb\ is completely dominated by the shocks.\ At low values of  \oiii/\oii, on the other hand, the effect of the shocks on  \oiii/\oii\ is imperceptible.

   \begin{figure*} 
   \centering
  \makebox[\textwidth][c]{\includegraphics[scale=0.45, trim={10 05mm 0 0mm}, clip]{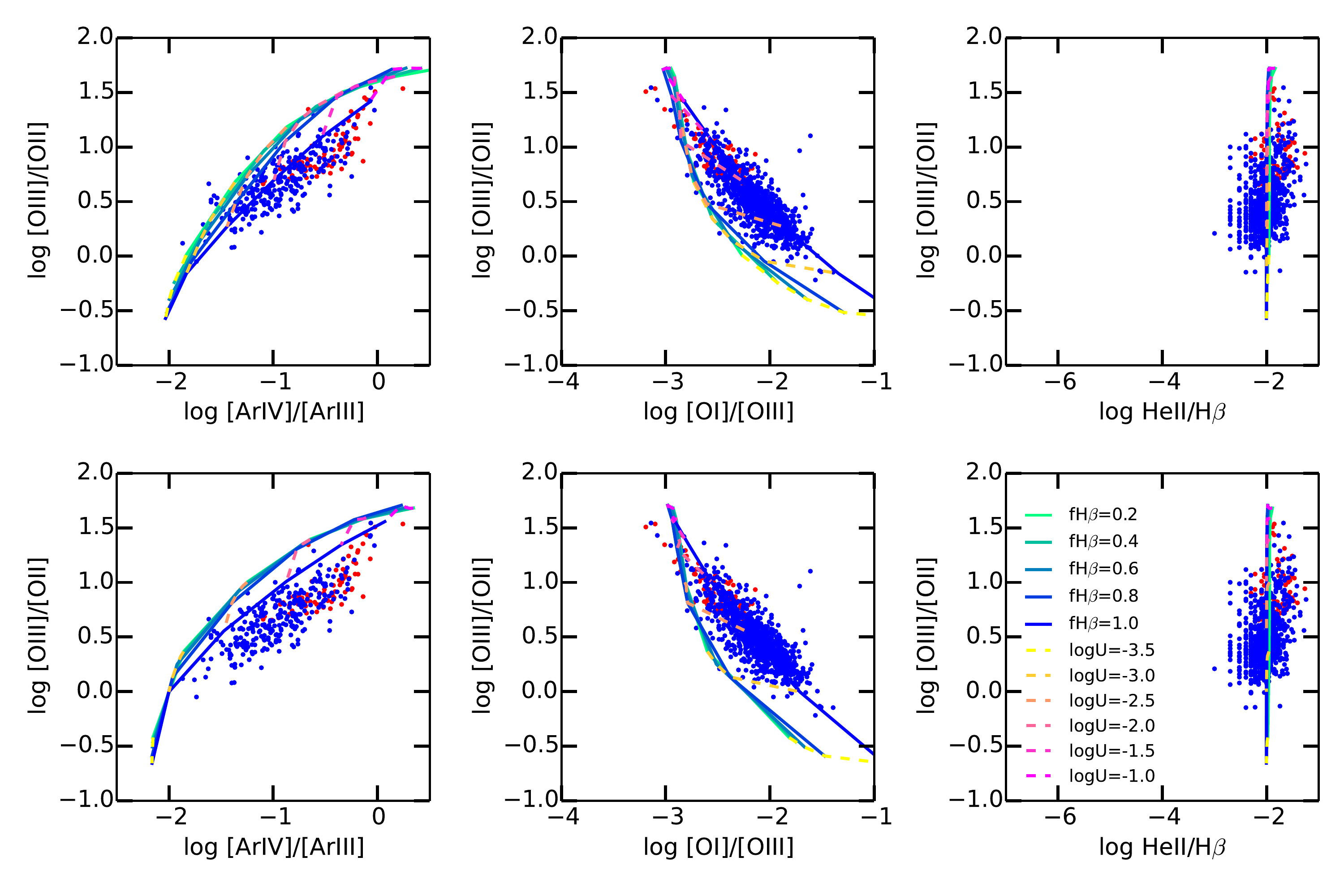}
}
      \caption{Same as Fig. \ref{fig4c}, but for shock models with a magnetic field of 10 $\mu$G.}
         \label{fig4c10}
   \end{figure*}

In Fig. \ref{fig4c10} we show the combination of the photoionization models using {\sc PopStar} with a shock model with a velocity of
300 km s$^{-1}$ and  a magnetic field strength of 10 $\mu$G, this time maximizing the contribution to \heii/\Hb\ and minimizing the contribution to \oi/\Hb. The value of $C$ to obtain \heii/\Hb\ $=0.01$  is   0.055. In this case,  the  grid of models follows the observational points in the  \oiii/\oii\ versus \oi/\oiii\ plot rather well, with however a slight deviation from the observed  trend at the highest values of \oiii/\oii.

We thus find that  for some combinations, a fraction of the observations  are in fact compatible with density-bounded models, implying a leakage of ionizing photons in these cases (but the \ariv/\ariii\ ratio remains unsatisfactory). Note that  at the highest values of \oiii/\oii\ the \oi/\oiii\ ratio is entirely determined by the shock and that the curves corresponding to different values of $f_{\Hb}$   merge. Estimating the amount of LyC photon leakage would at  least require a simultaneous fitting of all the relevant line ratios for each object and an independent estimate of the ionization parameter. This procedure would involve many unconstrained parameters, such as the shock velocities or the strength of the magnetic field. 

If, however, unlike what is assumed in our combined models, the shocks occur in the ionized region, the \oi\ emission would be different. 
From basic knowledge of the physics of ionized nebulae, one can intuit that the extreme UV and X-ray photons produced by the shock would travel until the outskirts of the \hii\ region and produce an  \oi\ emission similar to that of our {\sc PopStar} $+$ bremsstrahlung models.

From these considerations,  we conclude that the presence of shocks can enhance the  \heii/\Hb\ ratio and cause it  to agree with the observations and, at the same time,  enhance the \oi/\oiii\ ratio, especially for the objects of highest excitation. In these circumstances, an analysis based only on emission-line ratios cannot determine the existence or importance of LyC photon leakage.

IWe note, however, that  in the case of the objects with highest  excitation  the \oi/\oiii\ ratio is essentially determined by the shock. If the ionized nebulae in these objects were density-bounded, the tight correlation observed between  \oiii/\oii\ and \oi/\oiii\ would imply a very strong dependence of the shock properties on the characteristics of the photoionized nebula, in particular its ionization parameter. This is not impossible but is an additional requirement for the hypothesis of LyC photons leakage to be the correct interpretation.


\subsection{Models including the effect of an active nucleus}
\label{agn}

   \begin{figure*} 
   \centering
  \makebox[\textwidth][c]{\includegraphics[scale=0.45, trim={10 00mm 0 0mm}, clip]{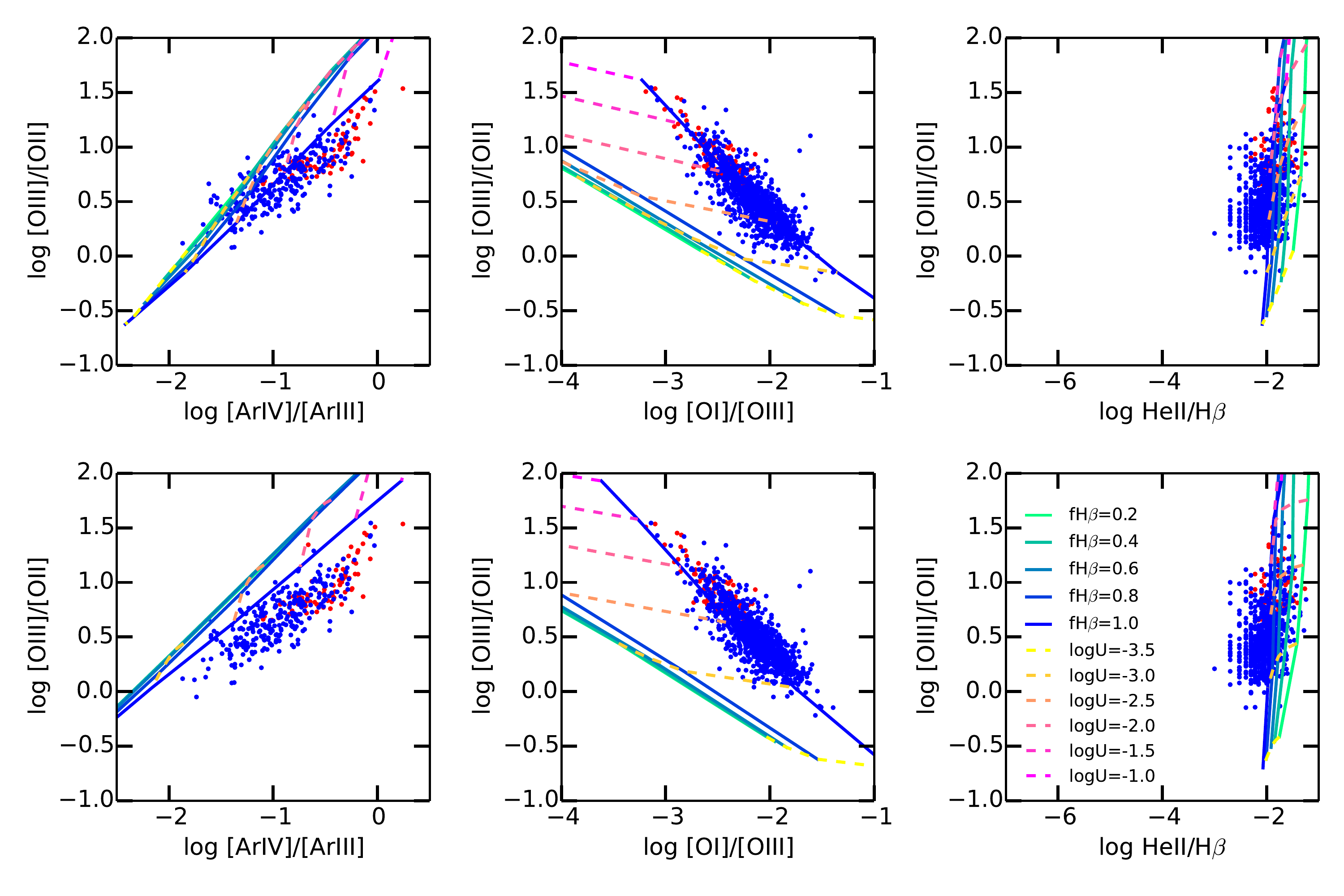}
}
      \caption{Same as Fig. \ref{figpopstar1}, but with models where a AGN-type ionizing spectrum  is added to the stellar ionizing radiation from the {\sc PopStar} models corresponding to an age of 1 Myr. The contribution of the AGN radiation is chosen to produce  a resulting \heii/\Hb\ of $\simeq 0.01$. 
      } 
         \label{figagn}
   \end{figure*}

The final scenario we consider is that of an AGN embedded in a star-forming region. Again, the SED of the stellar population is provided by the {\sc PopStar} models at an age of 1 Myr. For the AGN we took the standard AGN SED provided by {\sc Cloudy}.
 This AGN spectrum was combined with the {\sc PopStar} spectrum so as to obtain \heii/\Hb\ $\simeq 0.01$. The contribution of the AGN to the total number of ionizing photons must be quite small otherwise the \heii/\Hb\ ratio is boosted toward values much higher than observed (this contribution is about 3\% to obtain \heii/\Hb\ $ = 0.01$). The combined SED is shown in Fig. \ref{fig:seds_AGNbrem} of Appendix \ref{seds}.
The results of the photoionization models are shown in Fig. \ref{figagn}. Here, the models clearly indicate that the observed nebulae would be ionization-bounded. As in Figs. \ref{figpopstar1} and \ref{figbr}, the models give  \ariv/\ariii\ ratios somewhat lower than observed. 

\section{Summary and discussion}
\label{conclusions}

We have tested the hypothesis of LyC photon leakage from local extreme BCDs by assembling a sample of high-excitation galaxies from the SDSS DR 10 and DR7 and by examining their spectroscopic properties in the light of photoionization models. We have argued that a proper diagnostic cannot rely on strong lines alone: some weak lines are of prime importance to provide an unambiguous answer. Such are the \oi\ line, emitted in the warm transition region between ionized and neutral gas, and the \ariv\ and \heii\ lines, which are indicative of the presence of photons at high energies (40.8 eV and 54.4 eV, respectively) that might blur the information from the \oi\ lines.

After showing that photoionization models using SEDs from currently available stellar population synthesis models -- even at ages when Wolf-Rayet stars are present -- cannot account for all the observational constraints for the entire sample, we  considered several scenarios that could explain the data. These include the additional presence of a hard X-ray spectrum that might be produced by massive X-ray binaries or shocks in the atmospheres of stars  or protostars or by an active galactic nucleus. Another option would be  shocks that might be produced by supernovae or cloud-cloud collisions.  It turns out that only ionization-bounded models (or models with an escape fraction of Lyc photons lower than 10\%) are able to  explain all the observational constraints at the same time. 

We also found that the current stellar population synthesis models provide too soft ionizing radiation fields to explain the observed \ariv/\ariii\ ratios. This is probably because the stellar atmosphere models incorporated in the codes underpredict the number of photons emitted between 40.8 and 54.4 eV, and perhaps even at higher energies.

Our study   also showed that the extreme BCDs have higher ionization parameters (reaching log \mU  $=-1.5$ or even $-1$ depending on the nebular geometry) than galaxies with lower values of \oiii/\oii\ ratios. Such high values of the ionization parameters are probably produced by a more compact distribution of the gas around the ionizing stars.

The question of the possible leakage of LyC photons in high-redshift galaxies should be examined with methods similar to tthose presented in this paper, and not relying  on strong line ratios alone.  This will be possible with future deep surveys such as the Keck Baryonic Structure Survey (Steidel et al. 2014).

We must acknowledge, however,  that our study is based on idealized photoionization models. Perhaps some leakage might occur because of a nebular covering factor smaller than one and not because the nebulae are density-bounded? Perhaps the \oi\ emission actually comes from optically thick clumps embedded in a density-bounded medium?  Such hypotheses need to be tested by taking into account additional considerations (for example, imaging studies of nearby galaxies).  In the meantime,  our Occam's razor approach of photoionization modeling, the fact that  \oi/\oiii\ correlates so strongly with \oiii/\oii, and the fact that the highest \oiii/\oii\ ratios correlate with the highest  \Hb\ equivalent widths argue against strong LyC photon leakage from the galaxies with the highest observed \oiii/\oii\ ratios.  

\begin{acknowledgements}

We thank the referee for the interesting questions raised in his/her report, which prompted a deeper analysis on several points.

G. S. gratefully acknowledges financial support from the project UNAM PAPIIT IN109614 and the hospitality of the Instituto de Astronomia of the UNAM (Mexico), where part of this work was conducted. Yu. I. is grateful to the Observatoire de Paris for financial help and hospitality during the first stages of this work. C. M.  acknowledges financial aid from CONACyT project CB-2010/153985. N.G. acknowledges financial support by the Max Planck 
Institute for Radioastronomy in Bonn (MPIfR).

All the authors wish to thank the team of the
Sloan Digital Sky Survey (SDSS) for their dedication to a project
that has made the present work possible.

Funding for the SDSS and SDSS-II has been provided by the Alfred P. Sloan Foundation, the Participating Institutions, the National Science Foundation, the U.S. Department of Energy, the National Aeronautics and Space Administration, the Japanese Monbukagakusho, the Max Planck Society, and the Higher Education Funding Council for England. The SDSS Web Site is http://www.sdss.org/.

Funding for SDSS-III has been provided by the Alfred P. Sloan Foundation, the Participating Institutions, the National Science Foundation, and the U.S. Department of Energy Office of Science. The SDSS-III Web site is http://www.sdss3.org/.

The SDSS is managed by the Astrophysical Research Consortium for the Participating Institutions. The Participating Institutions are the American Museum of Natural History, Astrophysical Institute Potsdam, University of Basel, University of Cambridge, Case Western Reserve University, University of Chicago, Drexel University, Fermilab, the Institute for Advanced Study, the Japan Participation Group, Johns Hopkins University, the Joint Institute for Nuclear Astrophysics, the Kavli Institute for Particle Astrophysics and Cosmology, the Korean Scientist Group, the Chinese Academy of Sciences (LAMOST), Los Alamos National Laboratory, the Max-Planck-Institute for Astronomy (MPIA), the Max-Planck-Institute for Astrophysics (MPA), New Mexico State University, Ohio State University, University of Pittsburgh, University of Portsmouth, Princeton University, the United States Naval Observatory, and the University of Washington.

SDSS-III is managed by the Astrophysical Research Consortium for the Participating Institutions of the SDSS-III Collaboration including the University of Arizona, the Brazilian Participation Group, Brookhaven National Laboratory, Carnegie Mellon University, University of Florida, the French Participation Group, the German Participation Group, Harvard University, the Instituto de Astrofisica de Canarias, the Michigan State/Notre Dame/JINA Participation Group, Johns Hopkins University, Lawrence Berkeley National Laboratory, Max Planck Institute for Astrophysics, Max Planck Institute for Extraterrestrial Physics, New Mexico State University, New York University, Ohio State University, Pennsylvania State University, University of Portsmouth, Princeton University, the Spanish Participation Group, University of Tokyo, University of Utah, Vanderbilt University, University of Virginia, University of Washington, and Yale University.

\end{acknowledgements}


\begin{appendix}

\section{Chemical abundances}
\label{ab}

\begin{figure} [!hb]
\centering
  \makebox[\columnwidth][c]{\includegraphics[scale=0.95, trim={0 110mm 300 0mm}, clip]{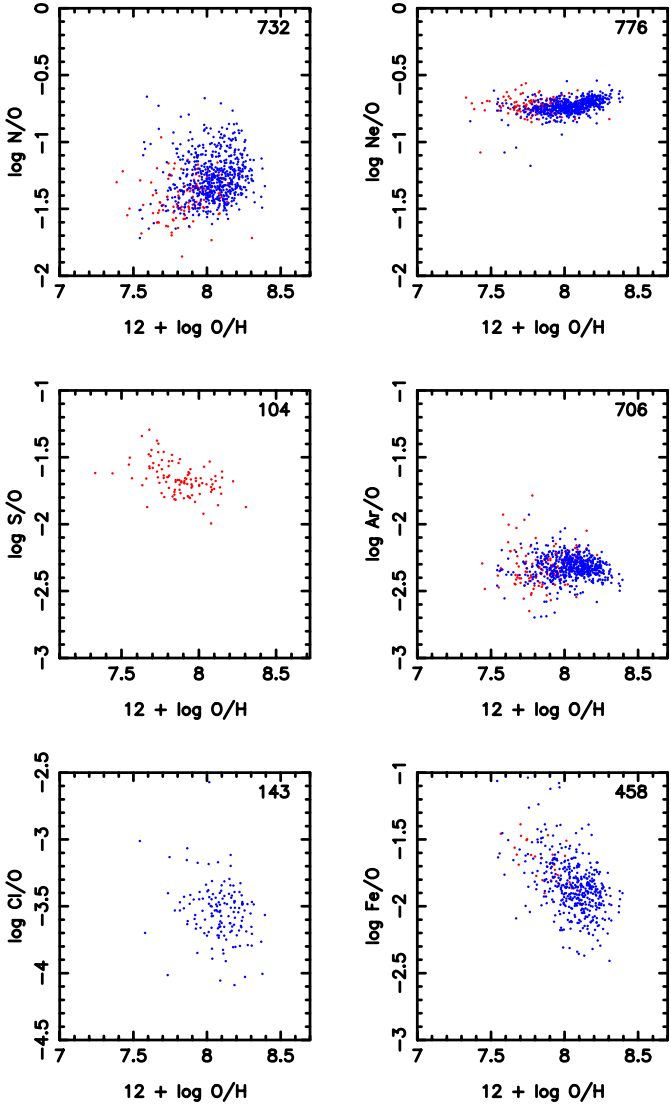}}
\caption{Abundance ratios in our galaxy sample. DR10 objects are plotted in red, DR7 objects are in blue. The total number of objects included in each panel is indicated in the upper right corner. }
 \label{figab}
\end{figure}

Figure \ref{figab} shows the abundance ratios N/O, Ne/O, S/O, Ar/O, Cl/O, Fe/O as a function of O/H for both our DR10 and DR7 samples (red and blue points, respectively). Note that we measured the S abundance only for objects from our DR10 sample since in the DR7 sample, the \Siii\ line for all the objects with measured \Oii\ lies beyond the spectroscopic range. The slight observed tendency between  S/O and O/H  might be due to inaccurate ionization correction factors. An improvement of the ionization correction factors presented in Izotov et al. (2006) would require a dedicated study that is beyond the scope of the present paper. The large dispersion in the computed Cl/O abundance ratios is essentially due to the extreme weakness of the lines and the large associated error bars. The dispersion in the N/O ratios  is real and discussed in detail in Vale Asari et al. (2014). The trend of Fe/O decreasing with increasing O/H, already noted by Izotov et al. (2006) using galaxies from SDSS DR3, is an indication that more Fe is depleted into dust grains as metallicity increases.

For the photoionization modeling described in Section \ref{photo} we took a chemical composition defined by the mean abundances of our DR7-DR10 sample. 


\section{Didactic diagrams}
\label{didactic}

   \begin{figure}  [!hbt]
   \centering
\makebox[\columnwidth][c]{\includegraphics[scale=0.65, trim={0 00mm 0 0mm}, clip]{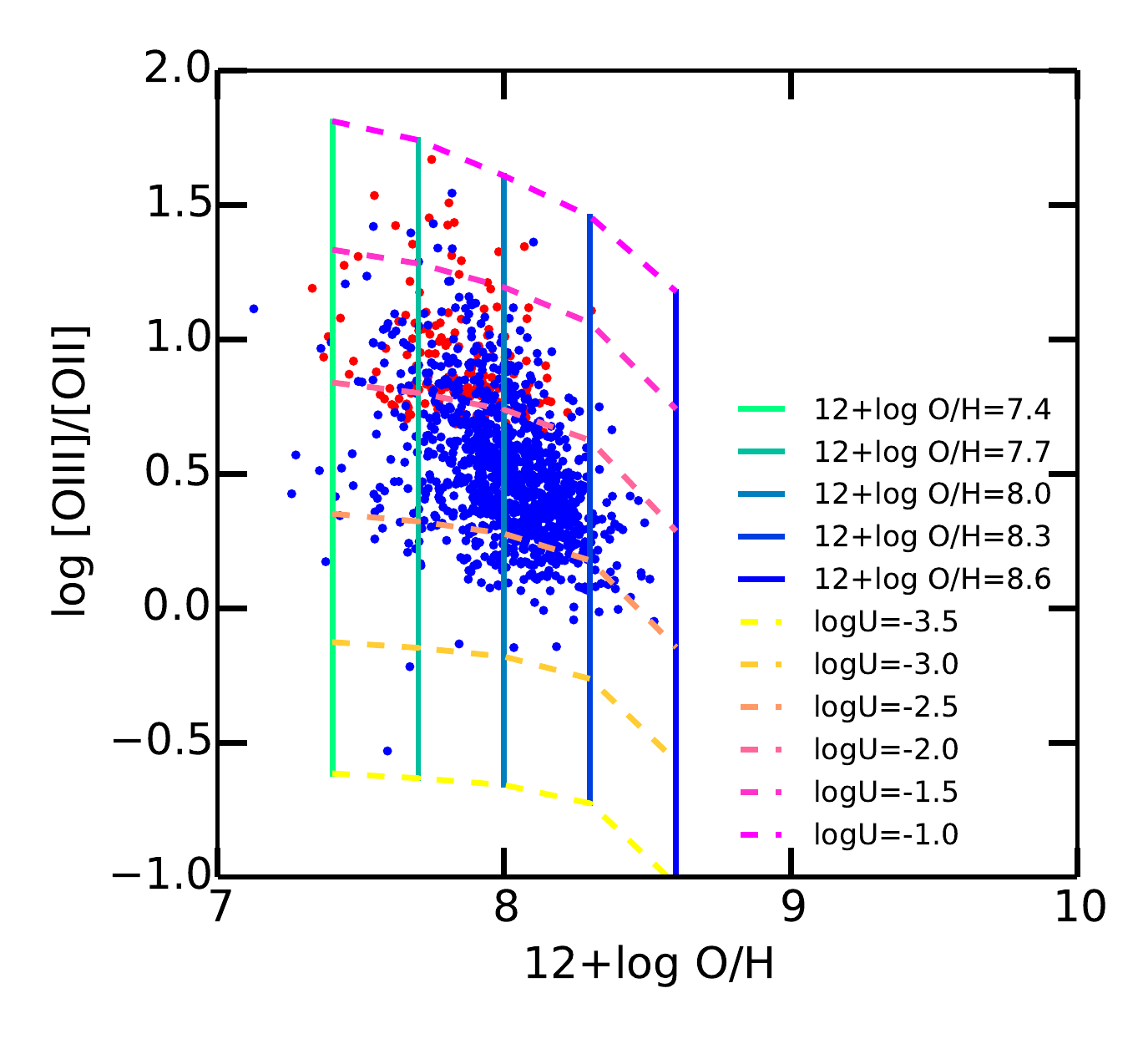}
}
      \caption{The effect of metallicity on the \oiii/\oii\ ratio obtained from photoionization models. The models are computed for a {\sc PopStar} ionizing radiation field (age 1 Myr) and are ionization-bounded. Continuous lines join models of same metallicity, while dashed lines join models with same ionization parameter. The color code is given in the plot. The positions of the observational points for our samples (red for DR10, blue for DR7) are shown for comparison.
      }
         \label{O3O2OsH}
   \end{figure}

   \begin{figure}  [!hbt]
   \centering
\makebox[\columnwidth][c]{\includegraphics[scale=0.45, trim={0 00mm 0 0mm}, clip]{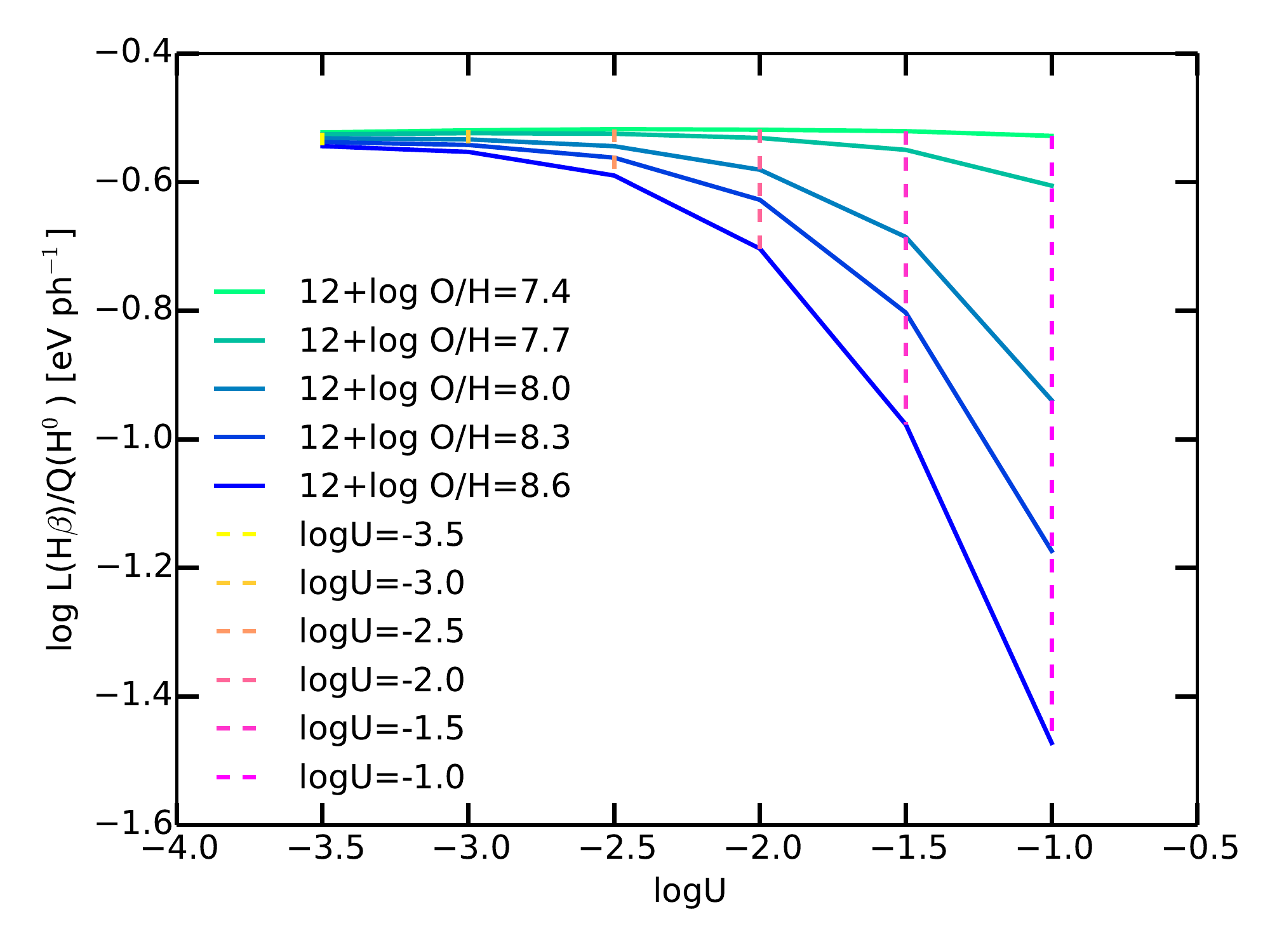}
}
      \caption{The values of $L(\Hb)/Q({\rm{H^{0}}})$ as a function of $\log \overline{U}_{\rm input}$ as computed in ionization-bounded photoionization models with different metallicities. The ionizing radiation field is from {\sc PopStar} at an age of 1 Myr. Continuous lines join models of same metallicity, while dashed lines join models with same ionization parameter. The color code is given in the plot. This diagram shows how the absorption of the ionizing radiation by dust grains affects the \Hb\ luminosity depending on the dust abundance (which is linked to the metallicity, see Sect. \ref{grid}) and on the ionization parameter.
      }
         \label{HbQ}
   \end{figure}

   \begin{figure}  [!h]
   \centering
  \makebox[\columnwidth][c]{\includegraphics[scale=0.47, trim={15 00mm 0 0mm}, clip]{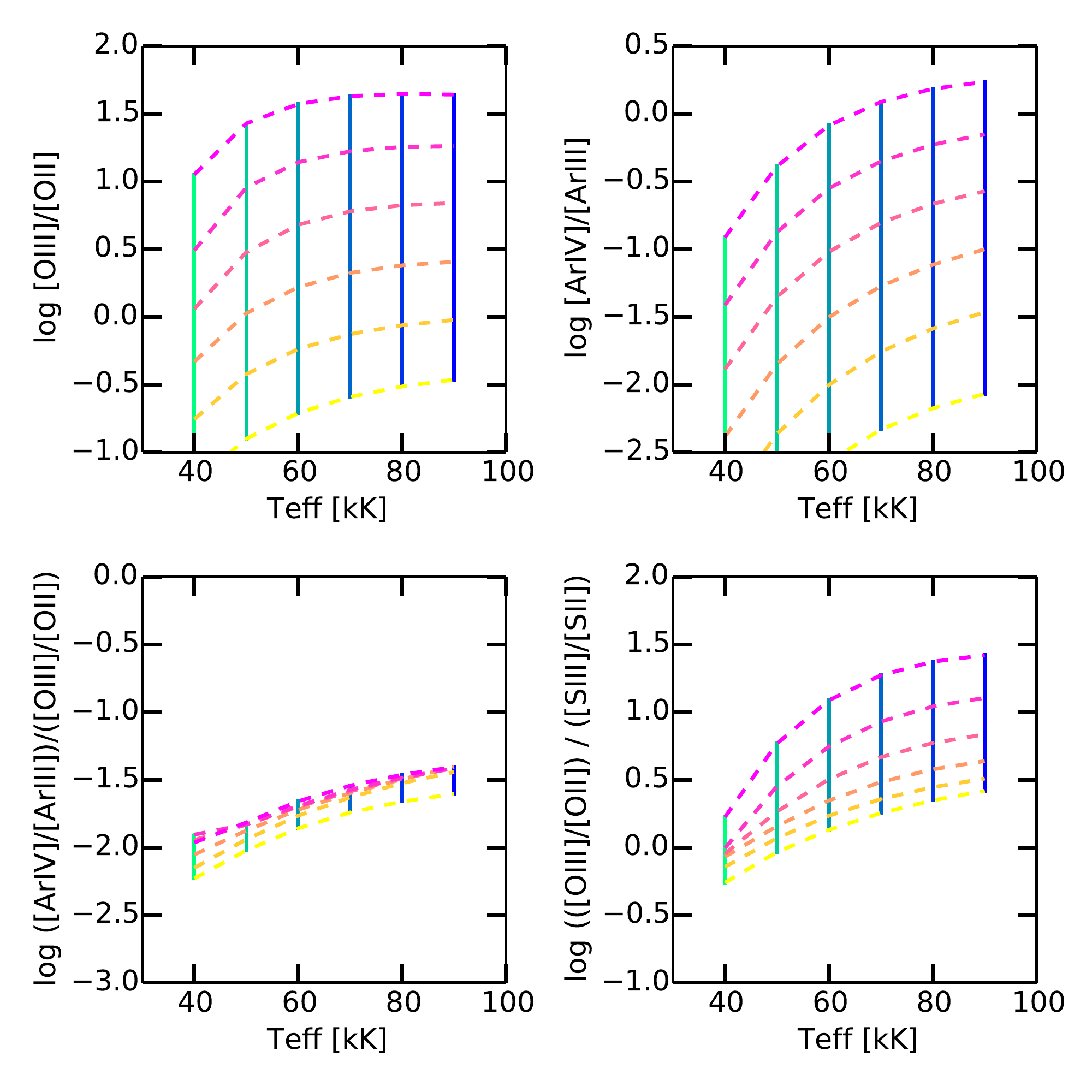}
}
      \caption{The variations of several line ratios with blackbody temperature for different values of the ionization parameter (ionization-bounded models with 12 + log O/H $=8$). Continuous lines join models of same effective temperature, while dashed lines join models with same ionization parameter. The color coding  is the same as in Fig. \ref{o3o2a4a3}.
      }
         \label{linerattiosvsT*}
   \end{figure}


\section{Spectral energy distributions}
\label{seds}

 
   \begin{figure} [!htbp]
   \centering
   \makebox[\columnwidth][c]{\includegraphics[scale=0.4, trim={0 00mm 0 0mm}, clip]{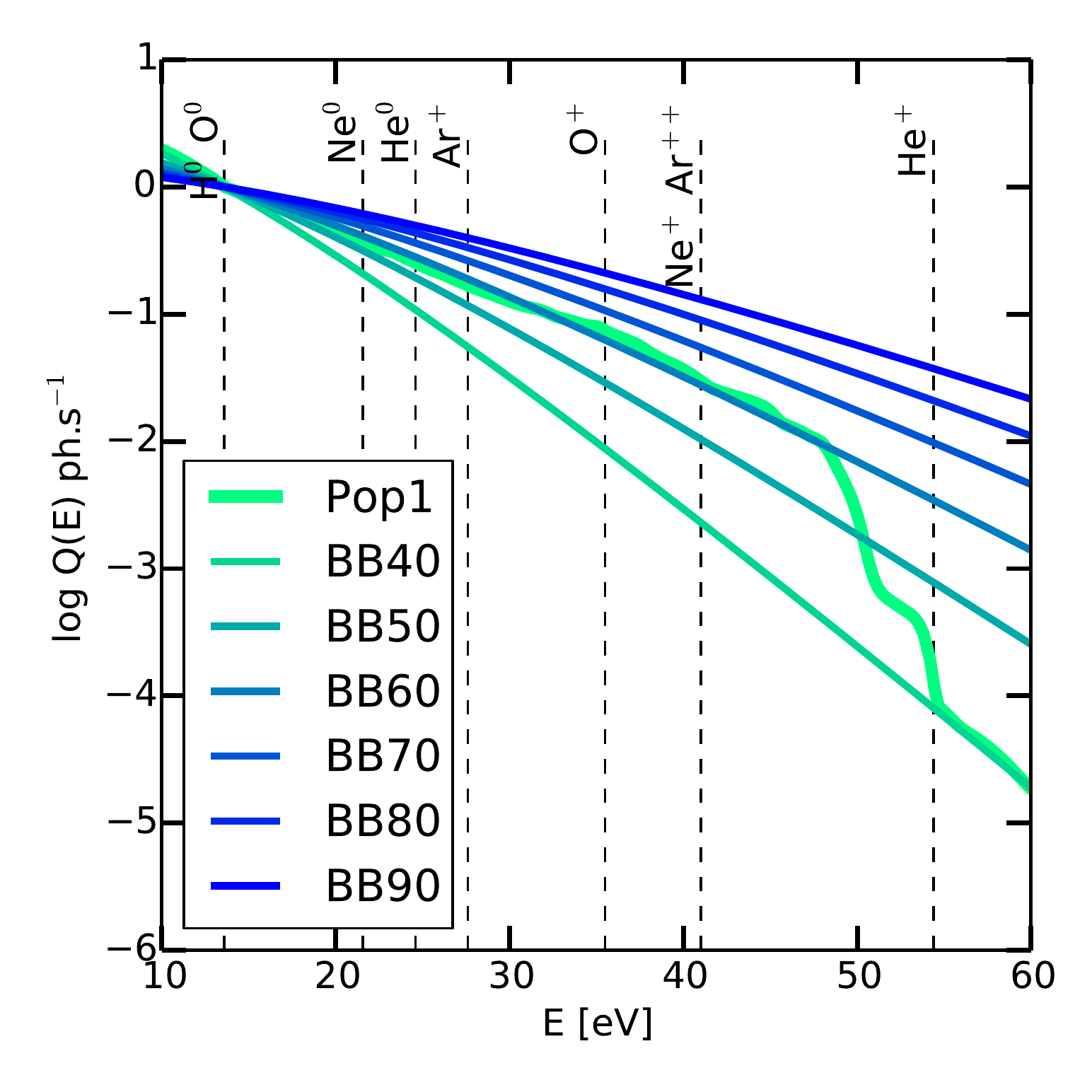}}
      \caption{The  log of $Q(E)$, the number of  photons emitted above the energy threshold $E$ as a function of $E$, for different spectral energy distributions (SEDs). The ionization thresholds of important ions are indicated by the vertical dashed lines. The different curves  correspond to blackbodies at different temperatures, as indicated in the inset. The {\sc PopStar} model at an age of 1 Myr of Fig. \ref{fig:seds_age} is shown for comparison.}
         \label{fig:seds_BB}
   \end{figure}

   \begin{figure} [!htbp]
   \centering
   \makebox[\columnwidth][c]{\includegraphics[scale=0.4, trim={0 00mm 0 0mm}, clip]{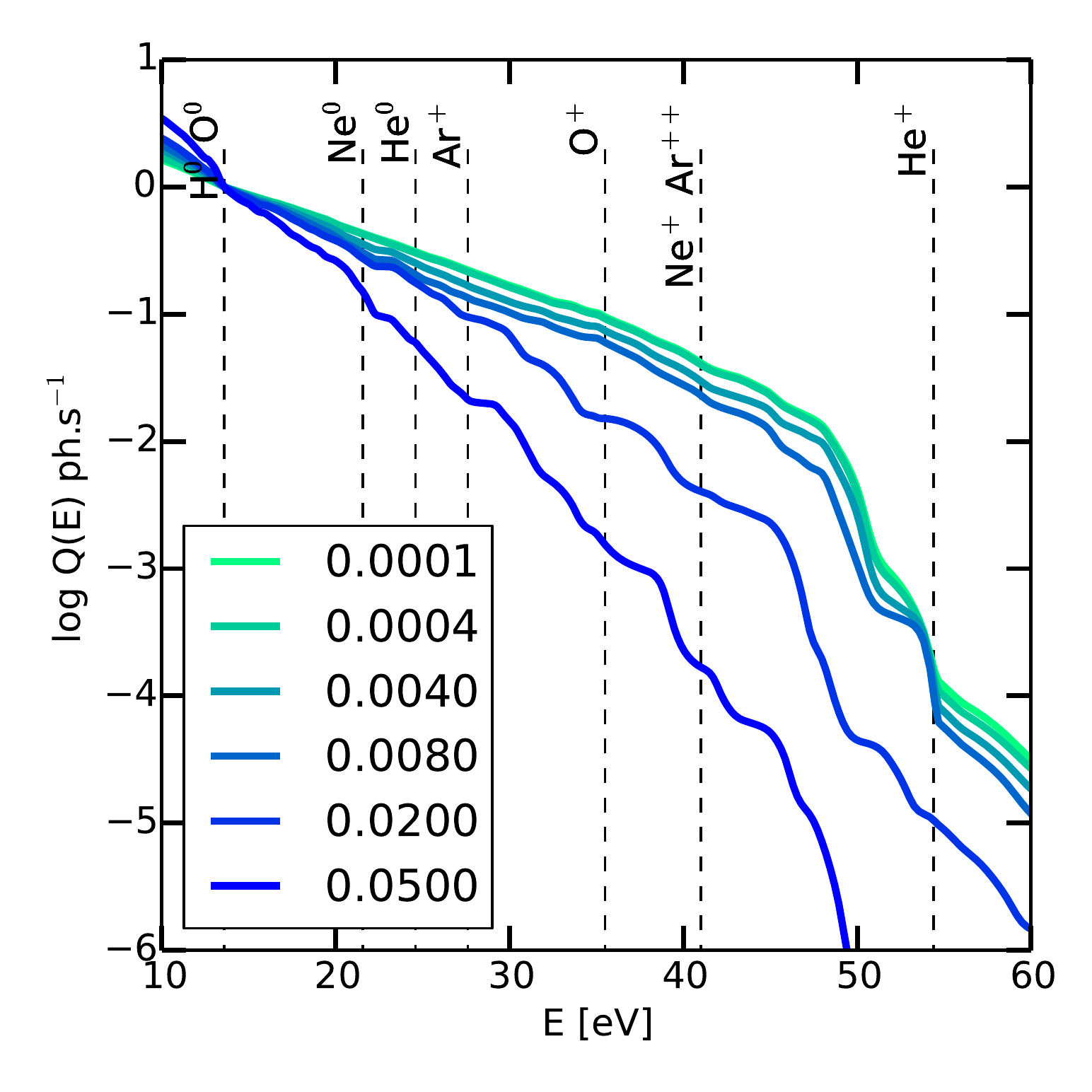}}
      \caption{Same as  Fig. \ref{fig:seds_BB} for SEDs obtained by {\sc PopStar}  at an age of 1 Myr for a Chabrier (2003) stellar initial mass and different metallicities  $Z$ as indicated by the color code in the inset.}
         \label{fig:SEDs_met}
   \end{figure}

   \begin{figure} [!htbp]
   \centering
   \makebox[\columnwidth][c]{\includegraphics[scale=0.4, trim={0 00mm 0 0mm}, clip]{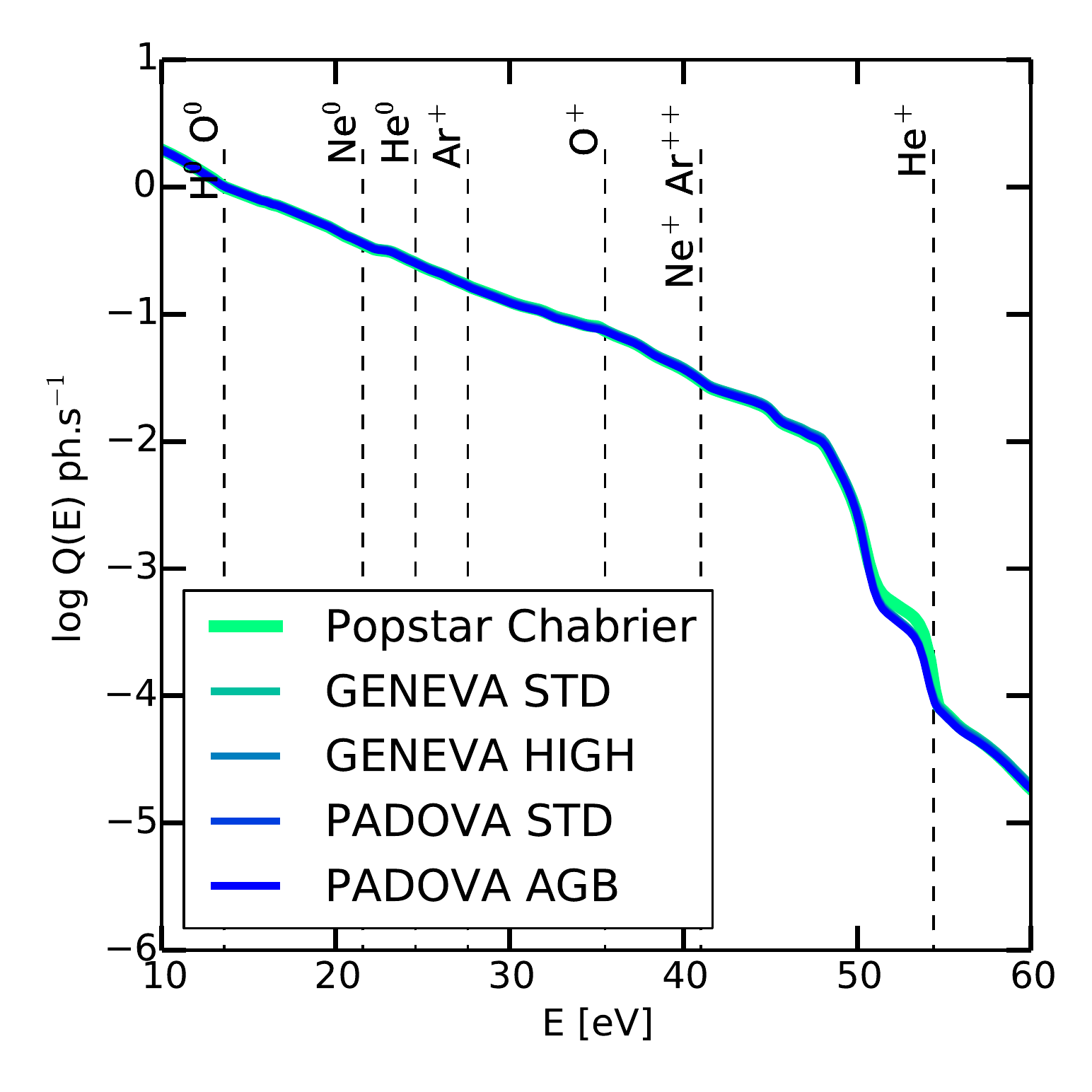}}
      \caption{Same as  Fig. \ref{fig:seds_BB} for several SEDs obtained with {\sc starburst99} with a Kroupa (2001) stellar initial mass function for a metallicity $Z = 0.004$ at an age of 1 Myr. The color coding indicates the different cases proposed by {\sc starburst99}: Geneva tracks with standard mass loss, Geneva tracks with high mass loss, original Padova tracks, and Padova tracks with AGB stars. The {\sc PopStar} model at an age of 1 Myr is shown for reference. We see that, at this age, all the SEDs are exactly sumperimposed.}
         \label{fig:seds_SEDs_1234_1}
   \end{figure}

   \begin{figure} [!htbp]
   \centering
   \makebox[\columnwidth][c]{\includegraphics[scale=0.4, trim={0 00mm 0 0mm}, clip]{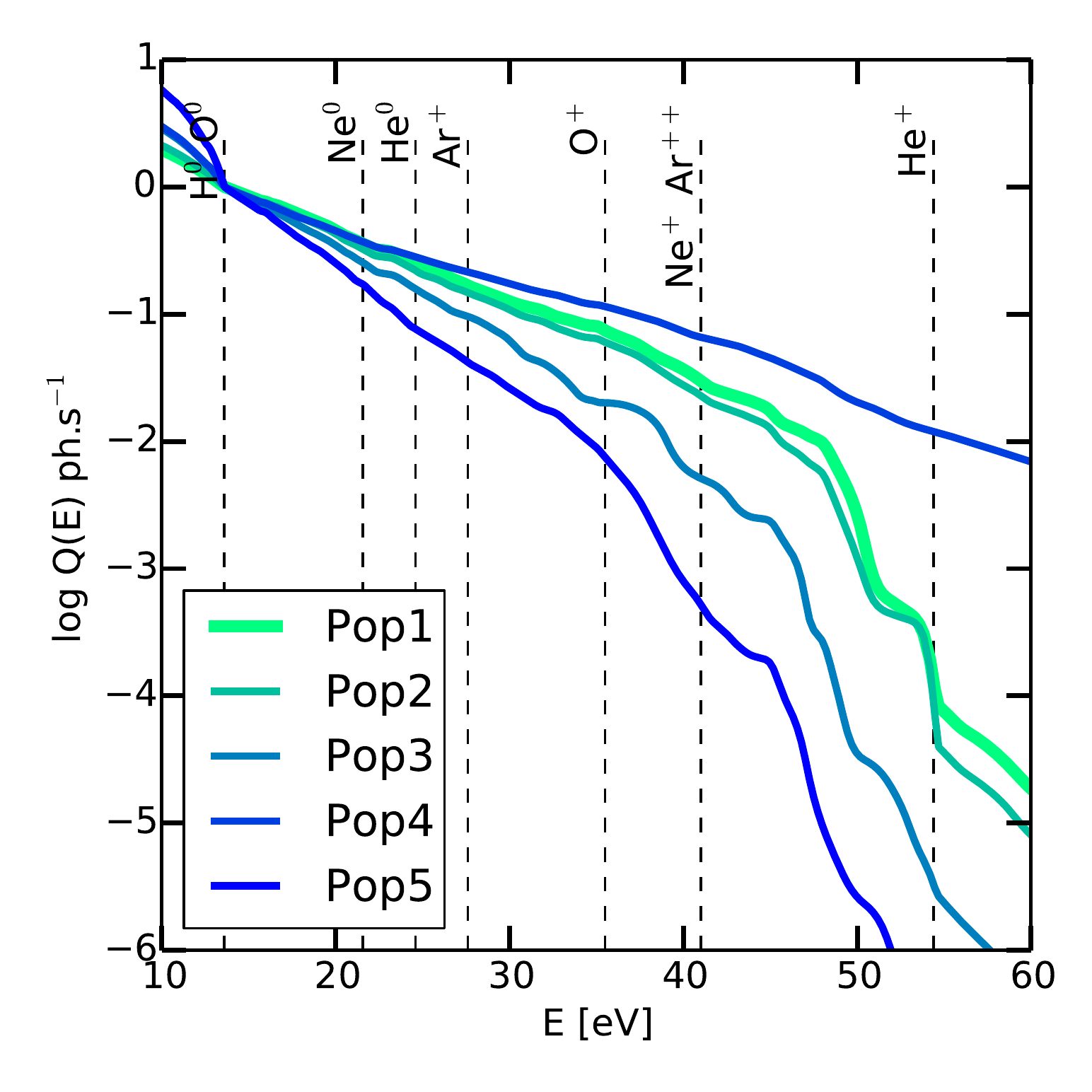}}
      \caption{Same as  Fig. \ref{fig:seds_BB}, but with SEDs corresponding
 to {\sc PopStar} models  at a metallicity $Z = 0.004$ for different ages (1, 2, 3, 4, and 5 Myr).}
         \label{fig:seds_age}
   \end{figure}

   \begin{figure} [!htbp]
   \centering
   \makebox[\columnwidth][c]{\includegraphics[scale=0.4, trim={0 00mm 0 0mm}, clip]{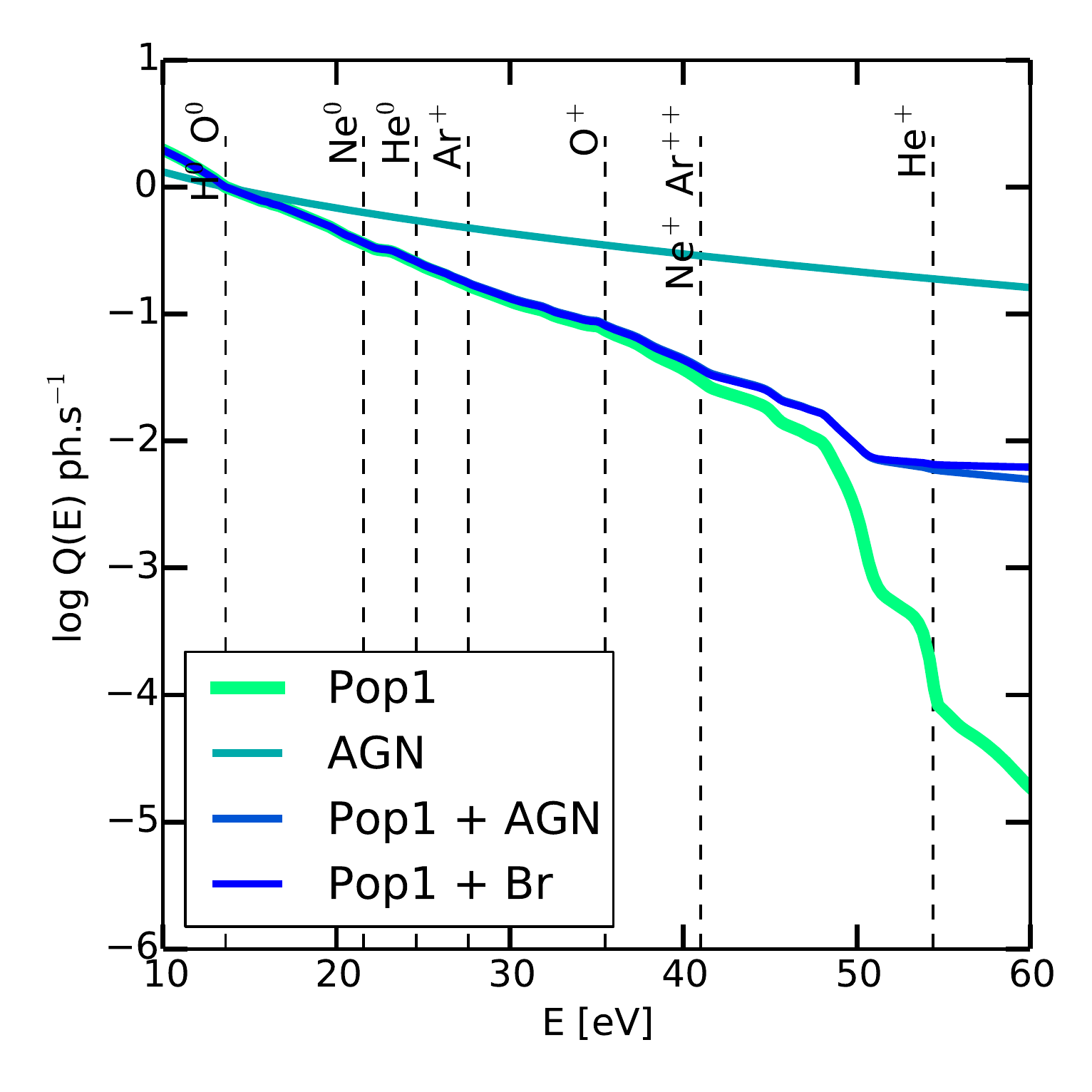}}
      \caption{Same as  Fig. \ref{fig:seds_age} for SEDs corresponding to AGN spectral energy distribution, combined {\sc PopStar} at an age of 1 Myr and AGN (Sect. \ref{agn}), and combined {\sc PopStar} at an age of 1 Myr  and bremsstrahlung (Sect. \ref{brem}). The pure {\sc PopStar} model at an age of 1 Myr of Fig. \ref{fig:seds_age} is shown for comparison.}
               \label{fig:seds_AGNbrem}
   \end{figure}

   \begin{figure} [!htbp]
   \centering
   \makebox[\columnwidth][c]{\includegraphics[scale=0.4, trim={0 00mm 0 0mm}, clip]{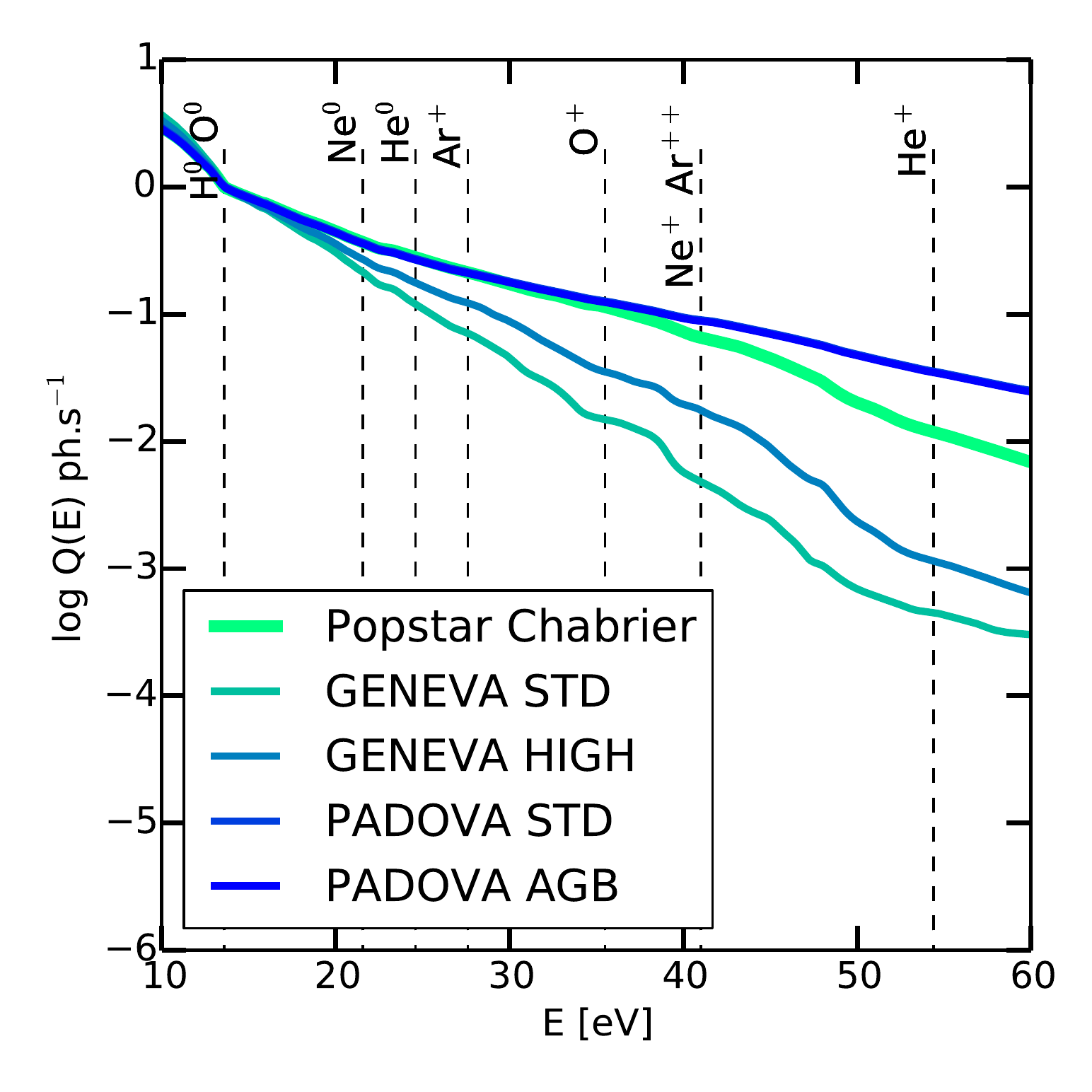}}
      \caption{Same as  Fig. \ref{fig:seds_SEDs_1234_1}, but for an age of 4 Myr. Here, all the SEDs are harder than in the 1 Myr case, because of   Wolf-Rayet stars. We also note a substantial difference between the various SEDs.}     
      \label{fig:seds_SEDs_1234_4}
   \end{figure}


\end{appendix}




\begin{thebibliography}{}

\bibitem[Abazajian et al.(2009)]{Abazajianetal2009}
{Abazajian}, K.~N., {Adelman-McCarthy}, J.~K., {Ag{\"u}eros}, M.~A. et
al., 2009, \apjs, 182, 543

\bibitem[\protect\citeauthoryear{Aggarwal 
\& Keenan}{1999}]{1999ApJS..123..311A} Aggarwal K.~M., Keenan F.~P., 1999, ApJS, 123, 311 


\bibitem[\protect\citeauthoryear{Ahn et al.}{2014}]{2014ApJS..211...17A} 
Ahn C.~P., et al., 2014, ApJS, 211, 17 

\bibitem[\protect\citeauthoryear{Allen et al.}{2008}]{2008ApJS..178...20A} 
Allen M.~G., Groves B.~A., Dopita M.~A., Sutherland R.~S., Kewley L.~J., 
2008, ApJS, 178, 20

\bibitem[Baldwin, Phillips and Terlevich(1981)]{BPT81}
{Baldwin}, J.~A. and {Phillips}, M.~M. and {Terlevich}, R.,1981, \pasp, 93, 5

\bibitem[\protect\citeauthoryear{Bonito et 
al.}{2010}]{2010A&A...517A..68B} Bonito R., Orlando S., Miceli M., Eisl{\"o}ffel J., Peres G., Favata F., 2010, A\&A, 517, AA68


\bibitem[\protect\citeauthoryear{Cardamone et 
al.}{2009}]{2009MNRAS.399.1191C} Cardamone C., et al., 2009, MNRAS, 399, 
1191 

\bibitem[\protect\citeauthoryear{Chabrier}{2003}]{2003ApJ...586L.133C} 
Chabrier G., 2003, ApJ, 586, L133 


\bibitem[\protect\citeauthoryear{Draine}{2011}]{2011ApJ...732..100D} Draine 
B.~T., 2011, ApJ, 732, 100

\bibitem[\protect\citeauthoryear{Ferland et 
al.}{2013}]{2013RMxAA..49..137F} Ferland G.~J., et al., 2013, RMxAA, 49, 
137

\bibitem[\protect\citeauthoryear{Froese Fischer, Tachiev, 
\& Irimia}{2006}]{2006ADNDT..92..607F} Froese Fischer C., Tachiev G., Irimia A., 2006, ADNDT, 92, 607 

\bibitem[\protect\citeauthoryear{Galavis, Mendoza, 
\& Zeippen}{1997}]{1997A&AS..123..159G} Galavis M.~E., Mendoza C., Zeippen C.~J., 1997, A\&AS, 123, 159 

\bibitem[\protect\citeauthoryear{Garnett et 
al.}{1991}]{1991ApJ...373..458G} Garnett D.~R., Kennicutt R.~C., Jr., Chu 
Y.-H., Skillman E.~D., 1991, ApJ, 373, 458



\bibitem[\protect\citeauthoryear{Hillier 
\& Miller}{1998}]{1998ApJ...496..407H} Hillier D.~J., Miller D.~L., 1998, ApJ, 496, 407 

\bibitem[\protect\citeauthoryear{Izotov, Guseva, 
\& Thuan}{2011}]{2011ApJ...728..161I} Izotov Y.~I., Guseva N.~G., Thuan T.~X., 2011, ApJ, 728, 161 

\bibitem[\protect\citeauthoryear{Izotov et 
al.}{2006}]{2006A&A...448..955I} Izotov Y.~I., Stasi{\'n}ska G., Meynet G., Guseva N.~G., Thuan T.~X., 2006, A\&A, 448, 955 


\bibitem[\protect\citeauthoryear{Izotov, Thuan, 
\& Privon}{2012}]{2012MNRAS.427.1229I} Izotov Y.~I., Thuan T.~X., Privon G., 2012, MNRAS, 427, 1229


\bibitem[\protect\citeauthoryear{Jaskot 
\& Oey}{2013}]{2013ApJ...766...91J} Jaskot A.~E., Oey M.~S., 2013, ApJ, 766, 91 

\bibitem[Kauffmann et al.(2003)]{Kauffmannetal2003}
{Kauffmann}, G. and {Heckman}, T.~M. and {Tremonti}, C. et al., 2003, \mnras, 346, 1055

\bibitem[\protect\citeauthoryear{Kewley et al.}{2013}]{2013ApJ...774L..10K} 
Kewley L.~J., Maier C., Yabe K., Ohta K., Akiyama M., Dopita M.~A., Yuan 
T., 2013, ApJ, 774, LL10 

\bibitem[\protect\citeauthoryear{Kisielius et 
al.}{2009}]{2009MNRAS.397..903K} Kisielius R., Storey P.~J., Ferland G.~J., 
Keenan F.~P., 2009, MNRAS, 397, 903 


\bibitem[\protect\citeauthoryear{Kroupa}{2001}]{2001MNRAS.322..231K} Kroupa 
P., 2001, MNRAS, 322, 231

\bibitem[\protect\citeauthoryear{Kudritzki}{2002}]{2002ApJ...577..389K} 
Kudritzki R.~P., 2002, ApJ, 577, 389 


\bibitem[\protect\citeauthoryear{Leitherer et 
al.}{1999}]{1999ApJS..123....3L} Leitherer C., et al., 1999, ApJS, 123, 3 

\bibitem[\protect\citeauthoryear{McLaughlin 
\& Bell}{2000}]{2000JPhB...33..597M} McLaughlin B.~M., Bell K.~L., 2000, JPhB, 33, 597 

\bibitem[\protect\citeauthoryear{Mendoza}{1983}]{1983IAUS..103..143M} 
Mendoza C., 1983, IAUS, 103, 143 

\bibitem[\protect\citeauthoryear{Mendoza \& Zeippen}{1983}]{1983MNRAS.202..981M} Mendoza C., Zeippen C.~J., 1983, MNRAS, 202, 981 

\bibitem[\protect\citeauthoryear{Moll{\'a}, Garc{\'{\i}}a-Vargas, 
\& Bressan}{2009}]{2009MNRAS.398..451M} Moll{\'a} M., Garc{\'{\i}}a-Vargas M.~L., Bressan A., 2009, MNRAS, 398, 451

\bibitem[\protect\citeauthoryear{Munoz Burgos et 
al.}{2009}]{2009A&A...500.1253M} Munoz Burgos J.~M., Loch S.~D., Ballance C.~P., Boivin R.~F., 2009, A\&A, 500, 1253 


\bibitem[\protect\citeauthoryear{Nakajima et 
al.}{2013}]{2013ApJ...769....3N} Nakajima K., Ouchi M., Shimasaku K., 
Hashimoto T., Ono Y., Lee J.~C., 2013, ApJ, 769, 3 


\bibitem[\protect\citeauthoryear{Nakajima 
\& Ouchi}{2014}]{2014MNRAS.442..900N} Nakajima K., Ouchi M., 2014, MNRAS, 442, 900

\bibitem[\protect\citeauthoryear{Overzier et 
al.}{2009}]{2009ApJ...706..203O} Overzier R.~A., et al., 2009, ApJ, 706, 
203 


\bibitem[\protect\citeauthoryear{Pauldrach, Hoffmann, 
\& Lennon}{2001}]{2001A&A...375..161P} Pauldrach A.~W.~A., Hoffmann T.~L., Lennon M., 2001, A\&A, 375, 161 

\bibitem[\protect\citeauthoryear{Ramsbottom, Bell, 
\& Keenan}{1997}]{1997MNRAS.284..754R} Ramsbottom C.~A., Bell K.~L., Keenan F.~P., 1997, MNRAS, 284, 754 


\bibitem[\protect\citeauthoryear{Ramsbottom, Bell, 
\& Keenan}{2001}]{2001ADNDT..77...57R} Ramsbottom C.~A., Bell K.~L., Keenan F.~P., 2001, ADNDT, 77, 57 


\bibitem[\protect\citeauthoryear{R{\'e}my-Ruyer et 
al.}{2014}]{2014A&A...563A..31R} R{\'e}my-Ruyer A., et al., 2014, A\&A, 563, AA31 


\bibitem[\protect\citeauthoryear{Rothschild et 
al.}{2013}]{2013ApJ...770...19R} Rothschild R., et al., 2013, ApJ, 770, 19 

\bibitem[\protect\citeauthoryear{Shapley et 
al.}{2014}]{2014arXiv1409.7071S} Shapley A.~E., et al., 2014, arXiv, 
arXiv:1409.7071 


\bibitem[\protect\citeauthoryear{Shirazi 
\& Brinchmann}{2012}]{2012MNRAS.421.1043S} Shirazi M., Brinchmann J., 2012, MNRAS, 421, 1043 


\bibitem[\protect\citeauthoryear{Shirazi, Brinchmann, 
\& Rahmati}{2014}]{2014ApJ...787..120S} Shirazi M., Brinchmann J., Rahmati A., 2014, ApJ, 787, 120


\bibitem[\protect\citeauthoryear{Smith, Norris, 
\& Crowther}{2002}]{2002MNRAS.337.1309S} Smith L.~J., Norris R.~P.~F., Crowther P.~A., 2002, MNRAS, 337, 1309 



\bibitem[\protect\citeauthoryear{Stasi{\'n}ska 
\& Izotov}{2003}]{2003A&A...397...71S} Stasi{\'n}ska G., Izotov Y., 2003, A\&A, 397, 71 



\bibitem[\protect\citeauthoryear{Stasi{\'n}ska 
\& Tylenda}{1986}]{1986A&A...155..137S} Stasi{\'n}ska G., Tylenda R., 1986, A\&A, 155, 137

\bibitem[\protect\citeauthoryear{Storey 
\& Zeippen}{2000}]{2000MNRAS.312..813S} Storey P.~J., Zeippen C.~J., 2000, MNRAS, 312, 813 


\bibitem[\protect\citeauthoryear{Tayal}{2011}]{2011ApJS..195...12T} Tayal 
S.~S., 2011, ApJS, 195, 12 


\bibitem[\protect\citeauthoryear{Tayal 
\& Gupta}{1999}]{1999ApJ...526..544T} Tayal S.~S., Gupta G.~P., 1999, ApJ, 526, 544 

\bibitem[\protect\citeauthoryear{Tayal 
\& Zatsarinny}{2010}]{2010ApJS..188...32T} Tayal S.~S., Zatsarinny O., 2010, ApJS, 188, 32 

\bibitem[\protect\citeauthoryear{valeasari}{}] VVale Asari et al., to be submitted

\bibitem[\protect\citeauthoryear{van Bever 
\& Vanbeveren}{2007}]{2007ASPC..367..579V} van Bever J., Vanbeveren D., 2007, ASPC, 367, 579 

\bibitem[\protect\citeauthoryear{Vilchez 
\& Pagel}{1988}]{1988MNRAS.231..257V} Vilchez J.~M., Pagel B.~E.~J., 1988, MNRAS, 231, 257

\bibitem[\protect\citeauthoryear{Zeippen}{1982}]{1982MNRAS.198..111Z} 
Zeippen C.~J., 1982, MNRAS, 198, 111

\end{thebibliography}
\end{document}